\documentclass[prd, tightenlines, superscriptaddress, showpacs, nofootinbib,
eqsecnum,amsfonts,amsmath]{revtex4}
\usepackage{bm,graphics,graphicx,epsfig,color,ulem} 
\allowdisplaybreaks
\def\be{\begin{equation}} 
\def\ee{\end{equation}} 
\def\ba{\begin{eqnarray}}
\def\ea{\end{eqnarray}} 
 
\def\lap{\lesssim}
 
\def\la{\langle} \def\ra{\rangle}
\def\bse{\begin{subequations}} 
\def\ese{\end{subequations}}
\def\rl{\right.\nonumber\\&\left.}
\def\rrll{\right.\right.\nonumber\\&\left.\left.}
\def\rrrlll{\right.\right.\right.\nonumber\\&\left.\left.\left.}
\def\rrrrllll{\right.\right.\right.\right.\nonumber\\&\left.\left.\left.\left.}
\def\rrrrrlllll{\right.\right.\right.\right.\right.\nonumber\\&\left.\left.\left.
\left.\left.}
 
\def\nn{\nonumber\\&} \def\p{\partial}

\begin{document}
\title{Third post-Newtonian gravitational waveforms for compact binary
systems in general orbits: Instantaneous terms}
\author{Chandra Kant Mishra}\email{chandra@icts.res.in}
\affiliation{Raman Research Institute, Bangalore 560 080, India}
\affiliation{Indian Institute of Science, Bangalore 560 012, India}
\affiliation{International Centre for Theoretical Sciences, Tata Institute 
of Fundamental Research, Bangalore 560012, India}
\author{K. G. Arun} \email{kgarun@cmi.ac.in} 
\affiliation{Chennai Mathematical Institute, Siruseri 603103, India}
\author{Bala R. Iyer} \email{bala.iyer@icts.res.in}
\affiliation{Raman Research Institute, Bangalore 560 080, India}
\begin{abstract}

We compute the {\it instantaneous} contributions to the spherical harmonic
modes of gravitational waveforms from compact binary systems in general orbits
up to the third post-Newtonian (PN) order. We further extend these results for
compact binaries in quasi-elliptical orbits using the 3PN quasi-Keplerian
representation of the conserved dynamics of compact binaries in eccentric
orbits. Using the multipolar post-Minkowskian formalism, starting from the
different mass and current-type multipole moments, we compute the spin-weighted
spherical harmonic decomposition of the instantaneous part of the gravitational
waveform. These are terms which are functions of the retarded time and do not
depend on the history of the binary evolution. Together with the {\it
hereditary} part, which depends on the binary's dynamical history, these
waveforms form the basis for construction of accurate templates for the
detection of gravitational wave signals from binaries moving in
quasi-elliptical orbits.

\end{abstract}
\date{\today}
\pacs{04.30.-w, 04.25.-g, 04.25.dg, 97.60.Jd}
\maketitle
\section{Introduction}
\label{ellWF:ellWFintro}

Compact binary systems composed of neutron stars and/or black holes are one of
the most promising sources for the second generation of earth-bound
gravitational-wave (GW) detectors such as Advanced LIGO
\cite{2014arXiv1411.4547T} and Advanced Virgo \cite{TheVirgo:2014hva} as well
as for the proposed space-based detector eLISA \cite{2012arXiv1201.3621A}.
Detection of such systems in GW detectors relies on a data-analysis technique
known as {\it matched filtering} which in turn requires very accurate modeling
of GW signals from these sources \cite{Th300}. The compact binaries are known
to have significant eccentricities when they are formed. However, since the GW
radiation reaction effects tend to circularize the binary's
orbit~\cite{PM63,Pe64}, for most long-lived binary systems one can expect that
their orbits would have circularized by the time they enter the sensitivity
band of ground-based detectors. This has motivated the GW community to perform
searches of GW signals from coalescing compact binary systems (CCBs) using
circular orbit templates.\\

Many astrophysical scenarios have been proposed which suggest the possible
existence of close eccentric binary systems.\footnote{Eccentric binary systems,
with small orbital separations, formed through the capture process. Unlike the
long-lived CCBs such systems are expected to enter the detector with nonzero
eccentricities.} One such scenario may exist in the cores of dense globular
clusters due to a mechanism known as the ``Kozai mechanism'' \cite{KozaiOsc62}.
This mechanism can also come into effect in scenarios involving formation of
hierarchical triples of supermassive black holes due to subsequent mergers of
galaxies \cite{BLSKozai02}. Another scenario might involve formation of {\it
close} eccentric compact binary systems in dense stellar systems like globular
clusters \cite{BenacquistaLiving}. Compact binaries involving intermediate mass
black holes in globular clusters might be seen in the eLISA band with residual
eccentricities of $0.1 \lap e \lap 0.2$ \cite{GMHimbh04}. Other scenarios
involve formation of close eccentric compact binary systems at centers of
galaxies \cite{Kupi:2006mh} and NS-BH binary systems which can become eccentric
as a consequence of multi stage mass transfer from the NS to the BH
\cite{DavLevKing04}. In light of these possibilities it becomes necessary to
compute accurate waveforms accounting for the eccentricity of the binary's
orbit.  \\

A number of investigations concerning the sensitivity of searches using
circular orbit templates to detect eccentric binary systems have been performed
in the past. The first such investigation was presented in
Ref.~\cite{PhysRevD.60.124008} where the authors studied the loss of
signal-to-noise ratio in detecting signals from binaries with residual
eccentricities using template banks constructed with quasi-circular waveform
models with leading order effects (both {\it conservative} and {\it secular}).
They argued that even if the system has a residual eccentricity ($e_0 \lap
0.13$ for binary system with two $1.4M_\odot$ neutron stars or $e_0 \lap 0.3$
for a binary with two $6M_\odot$ black holes), use of circular orbit templates
will be sufficient to detect signals from such systems.\footnote{They chose a
lower cutoff for the fitting factor (${\rm FF}\vert {\rm min}=90\%$)
corresponding to a loss in event rates of about less than $27\%$.} However,
this result has been subsequently weakened due to two independent
investigations \cite{PhysRevD.78.084029, PhysRevD.81.024007}.  Both of the
investigations suggest that if the eccentricity of the binary when it enters
the sensitivity band of detector is greater than 0.1, then it will not be
possible to detect such systems using circular orbit templates. These
investigations only dealt with sources that will be seen in ground-based
detectors. However, the capabilities of circular waveforms to detect signals
from the coalescence of supermassive black holes (visible in the eLISA
frequency band) have been investigated in \cite{Porter:2010mb}.  The results
presented in Ref.~\cite{Porter:2010mb} suggest that even to search signals from
sources with initial eccentricities of the order $10^{-4}$ one would need
waveforms which accurately account for the effects of eccentricities. In
addition, in a recent work, Huerta and Brown \cite{Huerta:2013qb} showed that
searches for CCBs with eccentricity $\geq 0.05$ would require eccentric
template banks to avoid significant loss in the sensitivity of the search.
Lastly, systematic errors due to the orbital eccentricity in measuring the
source parameters of double NS systems was investigated recently by
Favata~\cite{Favata2013} which again indicated the necessity to incorporate the
effects of eccentricity to measure the parameters of a double NS system if it
has non-negligible eccentricity when detected.\\
 
Evolution of a compact binary system can be divided into three stages: the
early inspiral, late inspiral and merger and the final ringdown. The early
inspiral phase can be very well modeled using the approximation schemes such
as multipolar post-Minkowskian (MPM) approximation matched to
post-Newtonian(PN)~\cite{Blanchet:2013haa} whereas the late inspiral, merger
and ringdown phases can be modeled using numerical relativity
(NR)~\cite{Pretorius07Review} or effective one body approach~\cite{BuonD98}. In
fact, it is now possible to perform numerical simulations to track the
evolution of the BH binary systems over many inspiral orbits and the subsequent
merger and ringdown phases.  However, {\it high} computational cost of
generating numerical waveforms covering the entire parameter space of
coalescing binary black holes (BBHs) has led to the construction of hybrid
waveforms (by combining PN and NR waveforms), which further are used to
phenomenologically construct a waveform model which has sufficient overlap with
the hybrid waveform \cite{ Pan:2011gk, Ajith:2007qp, Ajith:2007kx,
Ajith:2009bn, Buonanno:2007pf, Ajith:2012tt}. In addition to this, one needs to
check the consistency between these two waveforms (PN $\&$ NR) in a regime
where both of them are valid.  This would not only tell us about the
compatibility of the two waveforms but also would indicate the limits up to
which PN waveforms are reliable.  There have been many such investigations
involving nonspinning and nonprecessing BBH in quasi-circular orbits
\cite{BCCKM06, BCP07NR, NRPNGoddard07, NRPNJena07, Gopakumar:2007vh,
Boyle:2008ge, Damour:2007vq, Boyle:2009dg, Hannam:2010ec, Berti:2007fi,
NRPNCaltech07} and quasi-eccentric orbits \cite{2010PhRvD..82b4033H}. The need
for such comparisons and matching of the two waveforms (PN and NR) has led to
the high accuracy computations of spherical harmonic modes of the PN waveforms
in case of CCBs moving in quasi-circular orbits \cite{BFIS08, K08,
Faye:2012we}. Evidently, in order to  perform similar comparisons for eccentric
binaries, one would need high accuracy eccentric PN waveforms for such systems.
\\

The leading order (or Newtonian) expressions for the GW polarizations $(h_+
~{\rm and}~ h_\times)$ were obtained in the context of spacecraft Doppler
detection of GWs from an isolated compact binary in an eccentric orbit
\cite{1987GReGr..19.1101W}. This work was then extended to 1PN and the next
1.5PN order in \cite{WagW76, BS89, JunkS92, BS93, RS97}. At the 2PN order, the
transverse-traceless radiation field $(h^{\rm TT}_{ij})$ due to an isolated
binary composed of two compact stars moving in eccentric orbits was computed in
\cite{WWi96, GI97}.  Although the two works, \cite{WWi96} and \cite{GI97},
followed two different approaches, their final findings were in perfect
agreement with each other.  Under the {\it adiabatic} approximation, associated
2PN GW polarizations ($h_+, h_\times$) were obtained in \cite{GI02} for the
inspiral phase of binaries in quasi-eccentric orbits. Later in
Refs.~\cite{DGI04, KG06}, the method of variation of constants was used to
compute post-adiabatic corrections (varying on the orbital time scale and
$1/c^{5}$ times smaller) to the secular variation due to radiation reaction.
Using the 3PN generalized quasi-Keplerian representation of the conservative
dynamics of compact binary systems with arbitrary mass ratios moving in
eccentric orbits presented in \cite{MGS04}, Ref.~\cite{KG06} provides the
evolution of the orbital phase with relative 1PN accuracy (absolute 3.5PN).
The energy and angular momentum fluxes as well as evolution of orbital elements
up to 3PN order was calculated in Refs.~\cite{ABIQ07tail,ABIQ07,Arun:2009mc}.
Recently, computations of the frequency domain waveforms and the orbital
dynamics (both at the 2PN order) were presented for eccentric binaries in
harmonic coordinates \cite{Tessmer:2010ii}.  On the NR front, the first
simulations involving nonspinning equal mass BBHs in bound eccentric orbits
were performed in \cite{JenaEccentric07, HinderEccentric07} and the effects of
eccentricity on the final mass and spin were studied. Another recent work
\cite{2010PhRvD..82b4033H} presents numerical simulations for a nonspinning
equal mass binary system with an initial eccentricity of $e\sim 0.1$ and
compares the NR waveforms with those of the PN models.\\   

In this paper we present the computation of instantaneous part\footnote{The
part of the gravitational radiation which depends on the state of its source at
a given retarded time.} of various modes of the waveform ($h^{\ell m}$) for
general orbits using the basis of spherical harmonics of spin weight -2.  In
addition we also specialize to the case of compact binaries in quasi-elliptical
orbits and provide 3PN instantaneous expressions for various modes using 3PN
quasi-Keplerian representation of the conserved dynamics of compact binaries in
eccentric orbits \cite{MGS04, ABIQ07}.  Note again that investigations
presented here involve only the contributions from the instantaneous terms
which must be complemented by computations accounting for the hereditary
effects.\footnote{The part of the gravitational radiation which depends on the
entire dynamical history of the source and is complementary to the
instantaneous part of the radiation.} Computations of hereditary parts to
various modes of the waveform will form the basis for a companion paper
\cite{Mishra14}.\\

This paper is organized in the following manner. In
Sec.~\ref{ellWF:ellWFhlm-genformulas} we first introduce general formulas for
spherical harmonic modes of the gravitational waveform, $h^{\ell m}$, in terms
of the radiative mass and current multipole moments, $U^{\ell m}$ and $V^{\ell
m}$.  Section~\ref{ellWF:ellWFmoments} recalls some of the important aspects of
the MPM-PN formalism and lists various inputs that are needed for computing 3PN
expressions for various modes. These inputs involve relations connecting
radiative moments to source moments, expressions for various source multipole
moments for an isolated compact binary system and equations of motion.  In
Sec.~\ref{ellWF:ellWF3pnsphhar} we provide our results related to the
instantaneous part of the spherical harmonic modes of the waveform for a
nonspinning compact binary system in terms of variables that describe the
radiation from a generic compact binary. We find that these expressions are
quite large and run over several pages. Keeping this in mind we choose to list
only the dominant mode ($h_{22}$) in the main text of the paper and provide all
the relevant modes contributing to the 3PN waveform in a separate file readable
in MATHEMATICA (Hlm-GenOrb.m) that will be made available on the journal web
page as Supplemental Material \cite{SupplementalMaterial} along with the paper.
In Sec.~\ref{ellWF:ellWFQKR} we specialize to the case of CCBs moving in
quasi-elliptical orbits and provide the corresponding expression for the
dominant mode, $h^{22}$, in terms of the time-eccentricity $e_t$, a PN
parameter related to the orbital frequency $x$ and the eccentric anomaly $u$.
Similar to the general orbit case, in the case of CCBs in quasi-elliptical
orbits, expressions for all the relevant modes contributing to the 3PN waveform
will be listed in a separate file (Hlm-EllOrb.m). Finally in
Sec.~\ref{ellWF:ellWFconclusion} we conclude the paper by providing a brief
summary of our results and the future plans.

\section{Spherical harmonic modes of the gravitational
waveform} \label{ellWF:ellWFhlm-genformulas}
For an isolated source of GWs, the spherical harmonic modes of the waveform
($h^{\ell m}$), in terms of the radiative mass-type ($U^{\ell m}$) and
current-type multipole moments ($V^{\ell m}$) \cite{BFIS08, Th80, K08,
Faye:2012we}, are given as 
\begin{equation} \label{ellWF:inv} h^{\ell m} =
-\frac{G}{\sqrt{2}\,R\,c^{\ell+2}}\left[U^{\ell m}-\frac{i}{c}V^{\ell
m}\right], \end{equation}
where $R$ is the distance of the source in radiative coordinates, $G$ is
Newton's gravitational constant and $c$ is the speed of the light. The
radiative multipole moments, $U^{\ell m}$ and $V^{\ell m}$, appearing above are
related to the symmetric trace-free (STF) radiative moments $U_L$ and $V_L$ as
\begin{subequations}\label{ellWF:UV}\begin{align} U^{\ell m} &=
\frac{4}{\ell!}\,\sqrt{\frac{(\ell+1)(\ell+2)}{2\ell(\ell-1)}} \,\alpha_L^{\ell
m}\,U_L,\\ V^{\ell m} &=
-\frac{8}{\ell!}\,\sqrt{\frac{\ell(\ell+2)}{2(\ell+1)(\ell-1)}}
\,\alpha_L^{\ell m}\,V_L.  \end{align}\end{subequations}
Here $\alpha_L^{\ell m}$ denote STF tensors which connect the usual basis of
spherical harmonics $Y^{\ell m}(\Theta, \Phi)$ to the set of STF tensors
$N_{\langle L\rangle}=N_{\langle i_1}\cdots N_{i_\ell\rangle}$ as\footnote{Here
$L=i_1i_2\cdots i_l$ represents a multi-index composed of $l$ spatial indices
and the angular brackets ($\la \ra$) surrounding indices denote symmetric
trace-free projections.}
\begin{subequations}\label{ellWF:NY}\begin{align} N_{\langle
L\rangle}(\Theta,\Phi) &= \sum_{m=-\ell}^{\ell} \alpha_L^{\ell m}\,Y^{\ell
m}(\Theta,\Phi),\\Y^{\ell m}(\Theta,\Phi) &= \frac{(2\ell+1)!!}{4\pi \ell!}\,
\overline{\alpha}_L^{\ell m}\,N_{\langle L\rangle}(\Theta,\Phi).
\end{align}\end{subequations}

In the above, ${\bf N}={\bf X}/{\it R}$ is a unit vector pointing towards the
detector along the line joining the source to the detector. For instance, if
the binary's plane is the $x$-$y$ plane then ${\bf N}$, in terms of angles
($\Theta$, $\Phi$) giving the location of the binary, can be given as
\begin{align} {\bf N}&=\sin \Theta\,\cos \Phi\,\hat{\bf e}_x+\sin \Theta\,\sin
\Phi\,\hat{\bf e}_y +\cos \Theta \hat{\bf e}_z.  \end{align} 
The STF tensorial coefficients $\alpha_{L}^{\ell m}$ in terms of $N_{\langle
i_1}\cdots N_{i_\ell\rangle}$ and $Y^{\ell m}(\Theta, \Phi)$ can be written as
\footnote{The notation used in \cite{BFIS08, Faye:2012we} (which we follow
here) to the one in \cite{Th80,K08} is related by $\mathcal{Y}_L^{\ell
m}=\frac{(2\ell+1)!!}{4\pi \ell!} \, \overline{\alpha}_L^{\ell m}$.}
\begin{align}\label{ellWF:alpha} \alpha_L^{\ell m} &= \int d\Omega\,N_{\langle
L\rangle}\, \overline{Y}^{\,\ell m}\,,\end{align}
where the usual basis of spherical harmonics is given as
\begin{widetext} \begin{align} Y^{\ell m}(\Theta, \Phi)&=(-)^m\,{1 \over
2^\ell\,\ell!}\left[{2\,\ell+1 \over 4\,\pi} {(\ell-m)!
\over(\ell+m)!}\right]^{1/2}\mathrm{e}^{i\,m\,\Phi}(\sin \Theta)^m \,{{\rm
d}^{\ell+m} \over {\rm d}(\cos \Theta)^{\ell+m}}
\left(\cos^2\Theta-1\right)^\ell.  \end{align} \end{widetext}
It is important to note that for nonspinning binaries, there exists a mode
separation as pointed out in Ref.~\cite{K08} and explicitly shown in
Ref.~\cite{Faye:2012we}. The mode  $h^{\ell m}$ is completely determined by
mass-type radiative multipole moment ($U^{\ell m}$) when $\ell+m$ is even, and
by current-type radiative multipole moment ($V^{\ell m}$) when $\ell+m$ is odd.
This allows us to write for various modes 

\bse \label{ellWF:hlm-modesep} \begin{align} h^{\ell m} &=
-\frac{G}{\sqrt{2}\,R\,c^{\ell+2}}\,U^{\ell m} \quad\quad\quad\quad\quad {\rm
if}\,\ell+m\,\,{\rm is\, even},\\ h^{\ell m} &=
\frac{i\,G}{\sqrt{2}\,R\,c^{\ell+3}}\,V^{\ell m}
\,\,\,\,\,\quad\quad\quad\quad\quad {\rm if}\,\ell+m\,\,{\rm is\, odd}.
\end{align} \ese
\section{Inputs for computing the 3PN waveform} \label{ellWF:ellWFmoments}
\subsection{Relations connecting the radiative moments to the source 
moments}
\label{ellWF:ellWFmoments}

In the MPM-PN formalism \cite{Blanchet:2013haa, BFeom, BIJ02, BI04mult, BDE04,
BDEI05}, the radiative multipole moments ($U_{L}$, $V_{L}$) are first written
in terms of two sets of canonical moments ($M_{L}$, $S_{L}$), which in turn are
expressed in terms of six sets of source moments ($I_L$, $J_L$, $W_L$, $Y_L$,
$X_L$, $Z_L$). Relations connecting radiative moments to the canonical moments
and those connecting the canonical moments to the source moments, with the PN
accuracy desired for the waveform computations at the 3PN order, have been
established and have been listed in Ref.~\cite{BFIS08} (see Eqs.~(5.4)-(5.8) and
Eqs.~(5.9)-(5.11) there). Using these inputs we can parametrize the set of
radiative moments (and, hence, the modes) in terms of source multipole moments.
Below we list all the relevant radiative multipole moments in terms of the
source multipole moments with PN accuracy desired for the present work.
Further, these expressions can be decomposed into two parts namely the {\it
instantaneous} contribution and the {\it hereditary} contribution.\\  

The only radiative moment required at the 3PN order is the one related to the
mass quadrupole ($U_{ij}$) and is given by

\be U_{ij}=U_{ij}^{\rm inst}+U_{ij}^{\rm hered}, \label{ellWF:U2decom} \ee

where the instantaneous and hereditary parts in terms of the source multipole
moments read   

\begin{widetext} \bse\label{ellWF:U2} \begin{align}\label{ellWF:U2inst}
U_{ij}^{\rm inst}(U) &= I^{(2)}_{ij} (U) +{G \over c^5}\left\{{1
\over7}I^{(5)}_{a\langle i}I_{j\rangle a} - {5  \over7} I^{(4)}_{a\langle
i}I^{(1)}_{j\rangle a} -{2  \over7} I^{(3)}_{a\langle i}I^{(2)}_{j\rangle a}
+{1  \over3}\varepsilon_{ab\langle i}I^{(4)}_{j\rangle a}J_{b} +4\left[W^{(4)}
I_{ij}+W^{(3)} I_{ij}^{(1)}\rrll-W^{(2)} I_{ij}^{(2)}- W^{(1)}
I_{ij}^{(3)}\right]\right\} +\,\, \mathcal{O}\left(\frac{1}{c^7}\right),\\
U_{ij}^{\rm hered}(U) &= {2G M \over c^3} \int_{-\infty}^{U} d \tau \left[ \ln
\left({U-\tau \over 2\tau_0}\right)+{11 \over12} \right] I^{(4)}_{ij} (\tau)
+{G \over c^5}\left\{-{2 \over7}\int_{-\infty}^{U} d\tau I^{(3)}_{a\langle
i}(\tau)I^{(3)}_{j\rangle a}(\tau) \right\}\nonumber\\ & + 2\left({G M \over
c^3}\right)^2\int_{-\infty}^{U} d \tau \left[ \ln^2 \left({U-\tau
\over2\tau_0}\right)+{57 \over70} \ln\left({U-\tau \over
2\tau_0}\right)+{124627 \over44100} \right] I^{(5)}_{ij} (\tau) +\,\,
\mathcal{O}\left(\frac{1}{c^7}\right)\,.  \label{ellWF:U2hered}\end{align} \ese
\end{widetext}

In the above, the quantity $M$ represents the mass monopole moment or the
Arnowitt, Deser and Misner (ADM) mass of the source. The constant $\tau_0$
appearing in the above integrals is related to an arbitrary length scale $r_0$
by $\tau_0=r_0/c$ and was originally introduced in the MPM formalism. Note
that, numbers in the parenthesis (appearing as superscripts of the source
moments) denote the $p^\mathrm{th}$ time derivatives. The Levi-Civita tensor is
denoted by $\varepsilon_{ijk}$, such that $\varepsilon_{123}=+1$ and ${\cal
O}(1/c^7)$ indicates that we ignore contributions of order 3.5PN and
higher.\\

As may be seen from the above, computing the instantaneous part requires source
multipole moments given at a retarded time $U$. On the other hand, the
hereditary part involves integrals over time and would require the knowledge of
the source multipole moments at any instant of time before $U$ in the past
dynamical history of the source.   Further, the hereditary terms are of two
kinds: those with and without the logarithmic factors (see
Eq.~\eqref{ellWF:U2hered} above). The first integral appearing in
Eq.~\eqref{ellWF:U2hered} (the one with the logarithmic kernel inside) is
called the ``tail-integral'' and the one in the last line is called the
``tail-of-tail'' integral whereas the integral without the logarithmic factor
(in the first line) is known as the ``memory'' integral. This paper only
focuses on computing the {\it instantaneous} contribution to various modes of
gravitational waveforms and the computation of hereditary contributions shall
be discussed elsewhere \cite{Mishra14}.\\  

Moments required with 2.5PN accuracy are the mass octupole moment $U_{ijk}$ and
the current quadrupole moment $V_{ij}$. The mass octupole moment $U_{ijk}$ is given
as

\begin{align} U_{ijk}&=U_{ijk}^{\rm inst}+U_{ijk}^{\rm hered},
\label{ellWF:U3decom} \end{align}

where $U_{ij}^{\rm inst}$ and $U_{ijk}^{\rm hered}$ in terms of source
multipole moments read 

\begin{widetext} \begin{subequations}\label{ellWF:U3}\begin{align} U_{ijk}^{\rm
inst} (U) &= I^{(3)}_{ijk} (U) +{G \over c^5}\left\{ -{4
\over3}I^{(3)}_{a\langle i}I^{(3)}_{jk\rangle a}-{9 \over4}I^{(4)}_{a\langle
i}I^{(2)}_{jk\rangle a} + {1 \over4}I^{(2)}_{a\langle i}I^{(4)}_{jk\rangle a} -
{3 \over4}I^{(5)}_{a\langle i}I^{(1)}_{jk\rangle a} +{1
\over4}I^{(1)}_{a\langle i}I^{(5)}_{jk\rangle a}+ {1 \over12}I^{(6)}_{a\langle
i}I_{jk\rangle a}\rl+{1 \over4}I_{a\langle i}I^{(6)}_{jk\rangle a} + {1
\over5}\varepsilon_{ab\langle i}\left[-12J^{(2)}_{ja}I^{(3)}_{k\rangle
b}-8I^{(2)}_{ja}J^{(3)}_{k\rangle b} -3J^{(1)}_{ja}I^{(4)}_{k\rangle
b}-27I^{(1)}_{ja}J^{(4)}_{k\rangle b}-J_{ja}I^{(5)}_{k\rangle
b}-9I_{ja}J^{(5)}_{k\rangle b} \rrll-{9 \over4}J_{a}I^{(5)}_{jk\rangle
b}\right]+{12 \over5}J_{\langle i}J^{(4)}_{jk\rangle}+4\left[W^{(2)}
I_{ijk}-W^{(1)} I_{ijk}^{(1)} +3\, I_{\langle
ij}Y_{k\rangle}^{(1)}\right]^{(3)}\right\}
+\,\mathcal{O}\left(\frac{1}{c^6}\right) \label{ellWF:U3inst},\\
U_{ijk}^{\rm hered} (U) &={2G M \over c^3} \int_{-\infty}^{U} d\tau\left[ \ln
\left({U-\tau \over 2\tau_0}\right)+{97 \over60} \right] I^{(5)}_{ijk} (\tau)
+{G \over c^5}\left\{ \int_{-\infty}^{U} d\tau \left[-{1
\over3}I^{(3)}_{a\langle i} (\tau)I^{(4)}_{jk\rangle a} (\tau)\rrll-{4
\over5}\varepsilon_{ab\langle i} I^{(3)}_{ja} (\tau)J^{(3)}_{k\rangle b}
(\tau)\right]\right\} +\,\mathcal{O}\left(\frac{1}{c^6}\right)
\label{ellWF:U3hered}.  \end{align}\end{subequations}\end{widetext}

The current quadrupole moment $V_{ij}$ is given as

\begin{align} V_{ij}&=V_{ij}^{\rm inst}+V_{ij}^{\rm hered},
\label{ellWF:V2decom} \end{align}

where $V_{ij}^{\rm inst}$ and $V_{ij}^{\rm hered}$ in terms of source multipole
moments read 

\begin{widetext}\begin{subequations}\label{ellWF:V2}\begin{align} V_{ij}^{\rm
inst} (U) &= J^{(2)}_{ij} (U) + {G \over7\,c^{5}}\left\{4J^{(2)}_{a\langle
i}I^{(3)}_{j\rangle a}+8I^{(2)}_{a\langle i}J^{(3)}_{j\rangle a}
+17J^{(1)}_{a\langle i}I^{(4)}_{j\rangle a}-3I^{(1)}_{a\langle
i}J^{(4)}_{j\rangle a}+9I_{a\langle i}I^{(5)}_{j\rangle a} -3I_{a\langle
i}J^{(5)}_{j\rangle a}\rl-{1 \over4}J_{a}I^{(5)}_{ija} -7\varepsilon_{ab\langle
i}J_{a}J^{(4)}_{j\rangle b} +{1 \over2}\varepsilon_{ac\langle
i}\left[3I^{(3)}_{ab}I^{(3)}_{j\rangle bc} +{353 \over24}I^{(2)}_{j\rangle
bc}I^{(4)}_{ab} -{5 \over12}I^{(2)}_{ab}I^{(4)}_{j\rangle bc}+{113
\over8}I^{(1)}_{j\rangle bc}I^{(5)}_{ab} \rrll-{3
\over8}I^{(1)}_{ab}I^{(5)}_{j\rangle bc}+{15 \over4}I_{j\rangle bc}I^{(6)}_{ab}
+{3 \over8}I_{ab}I^{(6)}_{j\rangle bc}\right]+14\,\left[\varepsilon_{ab\langle
i}\left(-I_{j\rangle b}^{(3)}W_{a}-2I_{j\rangle b}Y_{a}^{(2)} +I_{j\rangle
b}^{(1)}Y_{a}^{(1)}\right)\rrll+3J_{\langle i}Y_{j\rangle
}^{(1)}-2J_{ij}^{(1)}W^{(1)}\right]^{(2)} \right\}
+\,\mathcal{O}\left(\frac{1}{c^6}\right)\,.\label{ellWF:V2inst}\\
V_{ij}^{\rm hered} (U) &= {2G M \over c^3} \int_{-\infty}^{U} d \tau \left[ \ln
\left({U-\tau \over 2\tau_0}\right)+{7 \over6} \right] J^{(4)}_{ij}(\tau)
+\,\mathcal{O}\left(\frac{1}{c^6}\right)\,.\label{ellWF:V2hered}
\end{align}\end{subequations}\end{widetext}

At the 2PN order the required moments are $U_{ijkl}$ and $V_{ijk}$. The moment
$U_{ijkl}$ is given by

\begin{align} U_{ijkl}&=U_{ijkl}^{\rm inst}+U_{ijkl}^{\rm hered},
\label{ellWF:U4decom} \end{align}

where $U_{ijkl}^{\rm inst}$ and $U_{ijkl}^{\rm hered}$ are related to the source
multipole moments by  

\begin{widetext}\begin{subequations}\label{ellWF:U4}\begin{align} U_{ijkl}^{\rm
inst}(U) &= I^{(4)}_{ijkl} (U) + {G \over c^3} \left\{ -{21
\over5}I^{(5)}_{\langle ij}I_{kl\rangle }
- {63 \over5}I^{(4)}_{\langle ij}I^{(1)}_{kl\rangle }- {102
  \over5}I^{(3)}_{\langle ij}I^{(2)}_{kl\rangle }\right\}
+\,\mathcal{O}\left(\frac{1}{c^5}\right)\label{ellWF:U4inst},\\
U_{ijkl}^{\rm hered} (U) &={G \over c^3} \left\{ 2 M \int_{-\infty}^{U} d \tau
\left[ \ln \left({U-\tau \over 2\tau_0}\right)+{59 \over30} \right]
I^{(6)}_{ijkl}(\tau)  +{2 \over5}\int_{-\infty}^{U} d\tau I^{(3)}_{\langle
ij}(\tau)I^{(3)}_{kl\rangle }(\tau) \right\}
+\,\mathcal{O}\left(\frac{1}{c^5}\right).\label{ellWF:U4hered}
\end{align}\end{subequations}\end{widetext}
The moment $V_{ijk}$ is given by 
\begin{align} V_{ijk}&=V_{ijk}^{\rm
inst}+V_{ijk}^{\rm hered}, \label{ellWF:V3decom} \end{align} 
where
$V_{ijk}^{\rm inst}$ and $V_{ijk}^{\rm hered}$ are given in terms of the source multipole
moments as  
\begin{widetext}\begin{subequations}\label{ellWF:V3}\begin{align}
V_{ijk}^{\rm inst} (U) &= J^{(3)}_{ijk} (U) + {G \over c^3} \left\{ {1
\over10}\varepsilon_{ab\langle i}I^{(5)}_{ja}I_{k\rangle b}- {1
\over2}\varepsilon_{ab\langle i}I^{(4)}_{ja}I^{(1)}_{k\rangle b} - 2 J_{\langle
i}I^{(4)}_{jk\rangle } \right\}
+\,\mathcal{O}\left(\frac{1}{c^5}\right),\label{ellWF:V3inst}\\ V_{ijk}^{\rm
hered} (U) &= {2 G M \over c^3}\int_{-\infty}^{U} d \tau \left[ \ln
\left({U-\tau \over 2\tau_0}\right)+{5 \over3} \right] J^{(5)}_{ijk} (\tau)
+\,\mathcal{O}\left(\frac{1}{c^5}\right).\label{ellWF:V3hered}
\end{align}\end{subequations}\end{widetext}

The moments required at the 1.5PN order are $U_{ijklm}$ and $V_{ijkl}$. The
mass-type moment $U_{ijklm}$ is given as 

\begin{align} U_{ijklm}&=U_{ijklm}^{\rm inst}+U_{ijklm}^{\rm hered},
\label{ellWF:U5decom} \end{align} where in terms of the source multipole moments, $U_{ijklm}^{\rm inst}$ and
$U_{ijklm}^{\rm hered}$ read
\begin{widetext}\begin{subequations}\label{ellWF:U5}\begin{align}
U_{ijklm}^{\rm inst}(U) &= I^{(5)}_{ijklm}(U) + {G \over c^3}\left\{ -{710
\over21}I^{(3)}_{\langle \ij}I^{(3)}_{klm\rangle} -{265 \over7}I^{(2)}_{\langle
ijk}I^{(4)}_{lm\rangle} -{120 \over7}I^{(2)}_{\langle ij}I^{(4)}_{klm\rangle}
-{155 \over7}I^{(1)}_{\langle ijk}I^{(5)}_{lm\rangle} \rl-{41
\over7}I^{(1)}_{\langle ij}I^{(5)}_{klm\rangle} -{34 \over7}I_{\langle
ijk}I^{(6)}_{lm\rangle} -{15 \over7}I_{\langle ij}I^{(6)}_{klm\rangle}\right\}
+\mathcal{O}\left(\frac{1}{c^4}\right)\label{ellWF:U5inst},\\
U_{ijklm}^{\rm hered}(U) &={G \over c^3}\left\{2 M \int_{-\infty}^{U} d \tau
\left[ \ln \left({U-\tau \over2\tau_0}\right)+ {232 \over105} \right]
I^{(7)}_{ijklm} (\tau)+{20 \over21}\int_{-\infty}^{U} d \tau I^{(3)}_{\langle
ij} (\tau) I^{(4)}_{klm\rangle} (\tau) \right\}\nonumber\\&
+\mathcal{O}\left(\frac{1}{c^4}\right)\label{ellWF:U5inst},
\end{align}\end{subequations}\end{widetext}
The current-type moment $V_{ijkl}$ is given by \begin{align}
V_{ijkl}&=V_{ijkl}^{\rm inst}+V_{ijkl}^{\rm hered}, \label{ellWF:V4decom}
\end{align}
where $V_{ijkl}^{\rm inst}$ and $V_{ijkl}^{\rm hered}$ in terms of the source
multipole moments read 
\begin{widetext}\begin{subequations}\label{ellWF:V4}\begin{align} V_{ijkl}^{\rm
inst}(U) &=J^{(4)}_{ijkl}(U) + {G \over c^3}\left\{-{35 \over3}S^{(2)}_{\langle
ij}I^{(3)}_{kl\rangle} -{25 \over3}I^{(2)}_{\langle ij}J^{(3)}_{kl\rangle} -{65
\over6}J^{(1)}_{\langle ij}I^{(4)}_{kl\rangle} -{25 \over6}I^{(1)}_{\langle
ij}J^{(4)}_{kl\rangle} -{19 \over6}J_{\langle ij}I^{(5)}_{kl\rangle}
\right.\nonumber\\ & -{11 \over6}I_{\langle ij}J^{(5)}_{kl\rangle} -{11
\over12}J_{\langle i}I^{(5)}_{jkl\rangle} +{1 \over6}\varepsilon_{ab\langle
i}\left[ -5I^{(3)}_{ja}I^{(3)}_{kl\rangle b} -{11
\over2}I^{(4)}_{ja}I^{(2)}_{kl\rangle b} -{5
\over2}I^{(2)}_{ja}I^{(4)}_{kl\rangle b} -{1
\over2}I^{(5)}_{ja}I^{(1)}_{kl\rangle b} \right.\nonumber\\&\left.\left.  +{37
\over10}I^{(1)}_{ja}I^{(5)}_{kl\rangle b} +{3 \over10}I^{(6)}_{ja}I_{kl\rangle
b} +{1 \over2}I_{ja}I^{(6)}_{kl\rangle b}\right] \right\}
+\mathcal{O}\left(\frac{1}{c^4}\right),\label{ellWF:V4inst}\\ 
V_{ijkl}^{\rm hered}(U) &= {2 G M \over c^3} \int_{-\infty}^{U} d \tau \left[
\ln \left({U-\tau \over2\tau_0}\right)+{119 \over60} \right]J^{(6)}_{ijkl}
(\tau)+\mathcal{O}\left(\frac{1}{c^4}\right)\,.\label{ellWF:V4hered}
\end{align}\end{subequations}\end{widetext}
Other mass-type moments $U_{L}$ contributing to 3PN waveform are given as
\be U_{L}=U_{L}^{\rm inst}+U_{L}^{\rm hered}, \label{ellWF:ULdecom} \ee where
$U_{L}^{\rm inst}$ and $U_{L}^{\rm hered}$ are related to the source multipole moments
as \begin{subequations}\label{ellWF:UL} \begin{align} U_{L}^{\rm inst} (U) &=
I^{(\ell)}_L(U) + \mathcal{O}\left(\frac{1}{c^3}\right)\,,\\ U_{L}^{\rm
hered}(U) &= \mathcal{O}\left(\frac{1}{c^3}\right)\,.
\end{align}\end{subequations}
Other current-type moments $V_{L}$ contributing to 3PN waveform are given as
\be V_{L}=V_{L}^{\rm inst}+V_{L}^{\rm hered}, \label{ellWF:VLdecom} \ee 
where finally $V_{L}^{\rm inst}$ and $V_{L}^{\rm hered}$ in terms of source multipole
moments read
\begin{subequations}\label{ellWF:VL} \begin{align} V_{L}^{\rm inst} (U) &=
J^{(\ell)}_L(U) + \mathcal{O}\left(\frac{1}{c^3}\right)\,,\\ V_{L}^{\rm
hered}(U) &= \mathcal{O}\left(\frac{1}{c^3}\right)\,.
\end{align}\end{subequations}

\subsection{Source multipole moments in general dynamical variables}
\label{ellWF:sourcemoms}

What we need next are expressions for various source multipole moments with the
PN accuracy sufficient for the present computation. Expressions for various
multipole moments presented here are generalizations of related circular orbit
expressions presented in \cite{ABIQ04, BFIS08} to the case of general orbits
and have been computed using the methods presented in \cite{BIJ02, BI04mult}.  We
skip all the details of the computation and list the final expressions for the
source multipole moments related to a source composed of two nonspinning
compact objects moving in general orbits.\\

The only moment required here with 3PN accuracy is the mass quadrupole,
$I_{ij}$, which for CCBs in general orbit was computed in Ref.~\cite{BI04mult} and
listed in Ref.~\cite{ABIQ07} in standard harmonic (SH)
coordinates.\footnote{Note that Ref.~\cite{ABIQ07} lists explicit expressions
for all the source multipole moments for binaries in general orbits needed for
computing 3PN energy flux.} As was argued in Ref.~\cite{ABIQ07}, though the use
of SH coordinate is useful in performing algebraic checks on PN computations,
quantities when expressed in these coordinates involve some {\it
gauge-dependent} logarithmic terms and are not suitable for numerical
calculations. It was suggested in Ref.~\cite{ABIQ07} that such logarithms can
be transformed away by using some coordinate transformations. They showed how
the use of a modified harmonic (MH) coordinate system (or alternatively an ADM
coordinate system) removes these logarithms.  We skip the details related to
those transformations and directly write the expression for the mass quadrupole
moment in MH coordinates. In MH coordinates, to 3PN accuracy, $I_{ij}$ reads 
 \begin{widetext} \begin{align} \label{ellWF:I2} I_{ij}&=\,\nu\,m\,\left\{
\left[A_1-\frac{24}{7}\,\frac{\,\nu}{c^5}\,\frac{G^2\,m^2}{r^2}\,\dot{r}\right]\,
x_{\la i}x_{j \ra}+\left[A_2\,\frac{r\,\dot{r}}{c^2}+\frac{48}{7}
\,\frac{\,\nu}{c^5}\,\frac{G^2\,m^2}{r}\right]\,x_{\la i}v_{j \ra}
+A_3\,\frac{r^2}{c^2}\,v_{\la i}v_{j\ra} \right\}+
\mathcal{O}\left(\frac{1}{c^7}\right), \end{align} \end{widetext}
where, \bse \label{ellWF:A1-A3} \begin{widetext} \begin{align} A_1&=1 +{1\over
c^2}\left[v^2 \left(\frac{29}{42}-\frac{29 \,\nu }{14}\right) +{G\,m\over r}
\left(-\frac{5}{7}+\frac{8 \,\nu }{7}\right)\right] +{1\over
c^4}\left[{G\,m\over r} v^2 \left(\frac{2021}{756}-\frac{5947 \,\nu }{756}
-\frac{4883 \,\nu ^2}{756}\right)\rl+{G\,m\over r} \dot{r}^2
\left(-\frac{131}{756} +\frac{907 \,\nu}{756}-\frac{1273 \,\nu ^2}{756}\right)
+{G^2\,m^2\over r^2} \left(-\frac{355}{252} -\frac{953 \,\nu }{126}+\frac{337
\,\nu ^2}{252}\right)+v^4\left(\frac{253}{504} -\frac{1835 \,\nu
}{504}\rrll+\frac{3545 \,\nu ^2}{504}\right)\right]
+{1\over c^6}\left[v^6 \left(\frac{4561}{11088}-\frac{7993 \,\nu }{1584}
+\frac{117067 \,\nu ^2}{5544}-\frac{328663 \,\nu ^3}{11088}\right)
+{G^2\,m^2\over r^2} \dot{r}^2\left(-\frac{8539}{20790}\rrll+\frac{52153 \,\nu
}{4158} -\frac{4652 \,\nu ^2}{231}-\frac{54121 \,\nu
^3}{5544}\right)+{G\,m\over r} \dot{r}^4 \left(\frac{2}{99}-\frac{1745 \,\nu
}{2772}+\frac{16319 \,\nu ^2}{5544} -\frac{311 \,\nu ^3}{99}\right)+{G\,m\over
r} v^4\,\times\rl\times \left(\frac{307}{77} -\frac{94475
\,\nu}{4158}+\frac{218411 \,\nu ^2}{8316}+\frac{299857 \,\nu ^3}{8316}\right)
+v^2 \left({G^2\,m^2\over r^2} \left(\frac{187183}{83160}-\frac{605419
\,\nu}{16632} +\frac{434909 \,\nu ^2}{16632}\rrrlll-\frac{37369 \,\nu
^3}{2772}\right) +{G\,m\over r} \dot{r}^2 \left(-\frac{757}{5544}+\frac{5545
\,\nu }{8316} -\frac{98311\,\nu ^2}{16632}+\frac{153407 \,\nu
^3}{8316}\right)\right) +{G^3\,m^3\over r^3}
\left(\frac{6285233}{207900}\rrll+\frac{15502 \,\nu }{385} -\frac{3632 \,\nu
^2}{693}+\frac{13289\,\nu ^3}{8316}-\frac{428}{105}
\log\left[\frac{r}{r_0}\right]\right)\right],\\
A_2&=-\frac{4}{7}+\frac{12 \,\nu }{7} +{1\over c^2}\left[v^2
\left(-\frac{26}{63}+\frac{202 \,\nu }{63} -\frac{418 \,\nu
^2}{63}\right)+{G\,m\over r} \left(-\frac{155}{54} +\frac{4057 \,\nu }{378}
+\frac{209 \,\nu ^2}{54}\right)\right] \nonumber\\&+{1\over
c^4}\left[{G\,m\over r} v^2 \left(-\frac{2839}{693} +\frac{237893 \,\nu
}{8316}-\frac{188063 \,\nu ^2}{4158} -\frac{58565 \,\nu ^3}{2079}\right)
+{G\,m\over r} \dot{r}^2 \left(\frac{305}{2772}+\frac{3233 \,\nu
}{2772}\rrll-\frac{8611 \,\nu ^2}{2772} -\frac{895 \,\nu
^3}{77}\right)+{G^2\,m^2\over r^2} \left(-\frac{12587}{20790} +\frac{406333
\,\nu}{8316}-\frac{2713 \,\nu ^2}{198}+\frac{4441 \,\nu ^3}{1386}\right) +v^4
\left(-\frac{457}{1386} \rrll+\frac{6103 \,\nu }{1386}-\frac{13693 \,\nu
^2}{693} +\frac{40687 \,\nu^3}{1386}\right)\right],\\
A_3&=\frac{11}{21}-\frac{11 \,\nu }{7} +{1\over c^2}\left[{G\,m\over r}
\left(\frac{106}{27}-\frac{335 \,\nu }{189} -\frac{985 \,\nu
^2}{189}\right)+\dot{r}^2 \left(\frac{5}{63}-\frac{25 \,\nu }{63} +\frac{25
\,\nu^2}{63}\right)+v^2 \left(\frac{41}{126}\rrll-\frac{337 \,\nu }{126}
+\frac{733 \,\nu ^2}{126}\right)\right]+{1\over c^4}\left[v^4
\left(\frac{1369}{5544} -\frac{19351 \,\nu }{5544}+\frac{45421 \,\nu
^2}{2772}-\frac{139999 \,\nu ^3}{5544}\right) +{G\,m\over r}
\dot{r}^2\left(\frac{79}{77} \rrll-\frac{5807 \,\nu }{1386} +\frac{515 \,\nu
^2}{1386}+\frac{8245 \,\nu ^3}{693}\right) +v^2 \left(\dot{r}^2
\left(\frac{115}{1386}-\frac{1135\,\nu }{1386} +\frac{1795 \,\nu
^2}{693}-\frac{3445 \,\nu ^3}{1386}\right) +{G\,m\over r} \left(\frac{587}{154}
\rrrlll-\frac{67933 \,\nu }{4158}+\frac{25660 \,\nu^2}{2079} +\frac{129781
\,\nu ^3}{4158}\right)\right)+{G^2\,m^2\over r^2} \left(-\frac{40716}{1925}
-\frac{10762 \,\nu }{2079}+\frac{62576 \,\nu ^2}{2079}-\frac{24314
\,\nu^3}{2079} \rrll+\frac{428}{105}
\log\left[\frac{r}{r_0}\right]\right)\right].  \end{align} \end{widetext} \ese
In the above, $x_i$ and $v_i$ denote the binary's relative separation and relative
velocity, respectively, whereas $\dot{r}$ denotes the radial velocity.  As we
see, the above expression still has a dependence on some logarithms
$(\log[r_0])$, where the quantity $r_0$ is related to the arbitrary constant
$\tau_0$ appearing in tail integrals by $\tau_0=r_0/c$. It has been argued and
shown that it disappears from all the physical quantities like the radiation field
at infinity and the far-zone energy flux~\cite{B98tail, BFIS08, K08, ABIQ07}.\\   

The expression for mass octupole, $I_{ijk}$, at 2.5PN order reads
\begin{widetext}\begin{align} \label{ellWF:I3}
I_{ijk}&=-\,\nu\,m\,\Delta\left\{\left[B_1-{56 \over9}{\,\nu \over
c^5}{G^2\,m^2 \over r^2} \,\dot{r}\right]\,x_{\la
ijk\ra}+\left[B_2\,{r\,\dot{r} \over c^2}+{\,\nu\,r \over c^5} \left({232
\over15}{G^2\,m^2 \over r^2}-{12 \over5}{G\,m \over r}\,v^2\right)\right]
\,x_{\la ij}v_{k\ra}\rl+B_3\,{r^2 \over c^2}\,x_{\la i}v_{jk\ra}
+B_4\,{r^3\,\dot{r} \over c^4}\,v_{\la ijk\ra}\right\}+
\mathcal{O}\left(\frac{1}{c^6}\right), \end{align}\end{widetext}
where,
\bse \begin{widetext} \label{ellWF:B1-B4} \begin{align}
B_1&=1+{1 \over c^2} \left[v^2 \left(\frac{5}{6}-\frac{19 \,\nu
}{6}\right)+{G\,m \over r}\left(-\frac{5}{6} +\frac{13 \,\nu
}{6}\right)\right]+{1 \over c^4}\left[{G\,m \over r} v^2 \left(
\frac{3853}{1320}-\frac{14257 \,\nu }{1320}-\frac{17371 \,\nu ^2}{1320}\right)
\rl+{G^2\,m^2 \over r^2} \left(-\frac{47}{33}-\frac{1591 \,\nu}{132} +\frac{235
\,\nu ^2}{66}\right)+v^4 \left(\frac{257}{440} -\frac{7319 \,\nu
}{1320}+\frac{5501 \,\nu ^2}{440}\right)+{G\,m \over r}\,\dot{r}^2
\left(-\frac{247}{1320}+\frac{531 \,\nu }{440}\rrll-\frac{1347 \,\nu
^2}{440}\right) \right],\\
B_2&=-\left(1-2 \,\nu\right)+{1 \over c^2}\left[v^2
\left(-\frac{13}{22}+\frac{107 \,\nu }{22} -\frac{102 \,\nu
^2}{11}\right)+{G\,m \over r} \left(-\frac{2461}{660} +\frac{8689 \,\nu
}{660}+\frac{1389 \,\nu^2}{220}\right)\right],\\
B_3&=1-2 \,\nu+{1 \over c^2}\left[{G\,m \over r}
\left(\frac{1949}{330}+\frac{62 \,\nu }{165} -\frac{483 \,\nu
^2}{55}\right)+v^2 \left(\frac{61}{110}-\frac{519 \,\nu }{110}+\frac{504 \,\nu
^2}{55}\right)\rl+\dot{r}^2\left(-\frac{1}{11}+\frac{4 \,\nu }{11}-\frac{3
\,\nu ^2}{11}\right) \right],\\
B_4&=\left(\frac{13}{55}-\frac{52 \,\nu }{55}+\frac{39 \,\nu ^2}{55}\right).
\end{align} \end{widetext} \ese
The remaining mass-type source multipole moments with PN accuracy required in
the present work are \bse \label{ellWF:I4-I8} \begin{widetext} \begin{align}
I_{ijkl}&=\,\nu\,m\left\{
x_{\la ijkl \ra}\left[1-3\,\nu+{1 \over c^2}\left[{G\,m \over r}
\left(-\frac{10}{11} +\frac{61 \,\nu }{11}-\frac{105 \,\nu ^2}{11}\right)+v^2
\left(\frac{103}{110} -\frac{147 \,\nu }{22}+\frac{279 \,\nu
^2}{22}\right)\right] \rrll+{1 \over c^4}\left[v^4
\left(\frac{3649}{5720}-\frac{50191 \,\nu }{5720} +\frac{112357 \,\nu
^2}{2860}-\frac{325687 \,\nu ^3}{5720}\right)+{G^2\,m^2 \over r^2}
\left(-\frac{15549}{10010}-\frac{9457 \,\nu }{715}+\frac{7961 \,\nu ^2}{143}
\rrrrllll-\frac{5829 \,\nu ^3}{286}\right)+{G\,m \over r} v^2
\left(\frac{11049}{3575} -\frac{152489\,\nu }{7150}+\frac{15124 \,\nu
^2}{715}+\frac{46934 \,\nu ^3}{715}\right) +{G\,m \over r}\,\dot{r}^2
\left(-\frac{659}{3575}\rrrrllll+\frac{12619 \,\nu }{7150} -\frac{10557
\,\nu^2}{1430}+\frac{9617 \,\nu ^3}{715}\right) \right] \right]
+x_{\la ijk}v_{l \ra}{r\,\dot{r} \over c^2}\left[-\frac{72}{55}+\frac{72 \,\nu
}{11} -\frac{72 \,\nu ^2}{11}+{1 \over c^2}\left[{G\,m \over
r}\,\times\rrrlll\,\times \left(-\frac{15463}{3575}+\frac{98374 \,\nu
}{3575}-\frac{25606 \,\nu ^2}{715} -\frac{18839 \,\nu ^3}{715}\right) +v^2
\left(-\frac{476}{715}+\frac{1228 \,\nu}{143}-\frac{23512 \,\nu ^2}{715}
\rrrrllll+\frac{25796 \,\nu ^3}{715}\right)\right] \right]
+x_{\la ij}v_{kl \ra}{r^2 \over c^2}\left[\frac{78}{55}-\frac{78 \,\nu }{11}
+\frac{78 \,\nu ^2}{11}+{1 \over c^2}\left[v^2 \left(\frac{553}{715}-\frac{6913
\,\nu }{715} +\frac{25994 \,\nu ^2}{715}\rrrrllll-\frac{28207 \,\nu
^3}{715}\right) +{G\,m \over r} \left(\frac{27818}{3575}-\frac{72474
\,\nu}{3575} -\frac{17202 \,\nu ^2}{715}+\frac{27568 \,\nu
^3}{715}\right)+\dot{r}^2\left(-\frac{4}{13} +\frac{28 \,\nu
}{13}\rrrrllll-\frac{56 \,\nu ^2}{13}+\frac{28 \,\nu ^3}{13}\right) \right]
\right]
+x_{\la i}v_{jkl \ra}{r^3\,\dot{r} \over c^4}\left[\frac{304}{715}-\frac{2128
\,\nu }{715} +\frac{4256 \,\nu ^2}{715}-\frac{2128 \,\nu ^3}{715} \right]
+v_{\la ijkl \ra}{r^4 \over
c^4}\,\times\rl\,\,\times\left[\frac{71}{715}-\frac{497 \,\nu }{715}+\frac{994
\,\nu ^2} {715}-\frac{497 \,\nu ^3}{715} \right] \right\}+
\mathcal{O}\left(\frac{1}{c^5}\right),\label{ellWF:I4}\\
I_{ijklm}&=-\,\nu\,m\,\Delta\left\{
x_{\la ijklm\ra}\left[1-2\,\nu+{1 \over c^2}\left[{G\,m \over r}
\left(-\frac{25}{26} +\frac{139 \,\nu }{26}-\frac{109 \,\nu ^2}{13}\right)+v^2
\left(\frac{79}{78} -\frac{511 \,\nu }{78}\rrrrllll+\frac{137 \,\nu
^2}{13}\right)\right]\right]
+x_{\la ijkl}v_{m\ra}{r\,\dot{r} \over c^2}\left[-\frac{20}{13}+\frac{80 \,\nu
}{13} -\frac{60 \,\nu ^2}{13}\right]
+x_{\la ijk}v_{lm\ra}{r^2 \over c^2}\left[\frac{70}{39}-\frac{280 \,\nu }{39}
+\frac{70 \,\nu ^2}{13}\right]
\right\}\nonumber\\&+\mathcal{O}\left(\frac{1}{c^4}\right),\label{ellWF:I5}\\
I_{ijklmn}&=\,\nu\,m\left\{
x_{\la ijklmn\ra}\left[1-5 \,\nu +5 \,\nu ^2+{1 \over c^2}\left[v^2
\left(\frac{15}{14} -\frac{21 \,\nu }{2}+33 \,\nu ^2-\frac{63 \,\nu
^3}{2}\right) -{G\,m \over r} \left(1-9 \,\nu \rrrrllll+27 \,\nu ^2-26 \,\nu
^3\right)\right]\right]
-x_{\la ijklm}v_{n\ra}\frac{12}{7}{r\,\dot{r} \over c^2}\left(1-7 \,\nu +14
\,\nu ^2 -7 \,\nu ^3\right)
+x_{\la ijkl}v_{mn\ra}\frac{15}{7}{r^2 \over c^2}\left(1-7 \,\nu \rrll+14 \,\nu
^2 -7 \,\nu ^3\right) \right\}+
\mathcal{O}\left(\frac{1}{c^4}\right),\label{ellWF:I6}\\
\label{ellWF:I7} I_{ijklmno}&=-\,\nu\,m\,\Delta\left(1-4 \,\nu +3 \,\nu
^2\right)x_{\la ijklmno\ra}+ \mathcal{O}\left(\frac{1}{c^2}\right),\\
\label{ellWF:I8} I_{ijklmnop}&=\,\nu\,m\,\left(1-7 \,\nu +14 \,\nu ^2-7 \,\nu
^3\right)x_{\la ijklmnop\ra}+ \mathcal{O}\left(\frac{1}{c^2}\right).
\end{align} \end{widetext} \ese
The current quadrupole moment is needed at 2.5PN order and given as
\begin{widetext} \begin{align} \label{ellWF:J2}
J_{ij}&=-\,\nu\,m\,\Delta\left\{\left[C_1 -{62 \over7}{{\dot{r}}\,\nu \over
c^5}{G^2\,m^2 \over r^2} \right]\epsilon_{ab\la i}x_{j\ra a}v_{b}\rl
+\left[C_2\,{r\,\dot{r} \over c^2} +{r\,\nu \over c^5}{G\,m \over r} \left({216
\over35}{G\,m \over r} -{4 \over5}\,v^2\right)\right]\epsilon_{ab\la i}v_{j\ra
b}x_{a}\right\}+ \mathcal{O}\left(\frac{1}{c^6}\right),
\end{align}\end{widetext}
where,
\bse \begin{widetext} \label{ellWF:C1-C2} \begin{align}
C_1&=1+{1 \over c^2}\left[v^2 \left(\frac{13}{28}-\frac{17 \,\nu }{7}\right)
+{G\,m \over r} \left(\frac{27}{14}+\frac{15 \,\nu }{7}\right)\right] +{1 \over
c^4}\left[{G\,m \over r} v^2 \left(\frac{671}{252}-\frac{1297 \,\nu }{126}
-\frac{121 \,\nu ^2}{12}\right)\rl+{G\,m \over r} \dot{r}^2
\left(-\frac{5}{252} -\frac{241 \,\nu}{252}-\frac{335 \,\nu ^2}{84}\right)
+{G^2\,m^2 \over r^2} \left(-\frac{43}{252}-\frac{1543 \,\nu }{126} +\frac{293
\,\nu ^2}{84}\right)+v^4 \left(\frac{29}{84}-\frac{11\,\nu }{3} \rrll+\frac{505
\,\nu ^2}{56}\right)\right],\\
C_2&=\frac{5}{28} (1-2 \,\nu )+\frac{1}{504} {1 \over c^2}\left[{G\,m \over r}
\left(824 +1348 \,\nu -1038 \,\nu ^2\right)+75 v^2 \left(1-7 \,\nu +12 \,\nu
^2\right)\right].  \end{align} \end{widetext} \ese
Other current-type source multipole moments with PN accuracies sufficient for
present calculations read \bse \label{ellWF:J3-J7} \begin{widetext}
\begin{align} \label{ellWF:J3} J_{ijk}&=\,\nu\,m\,\epsilon_{ab\la i}\left\{
x_{jk\ra a}v_{b}\left[1-3 \,\nu+{1 \over c^2}\left[{G\,m \over r}
\left(\frac{14}{9} -\frac{16 \,\nu }{9}-\frac{86 \,\nu ^2}{9}\right)+v^2
\left(\frac{41}{90}-\frac{77 \,\nu }{18} +\frac{185 \,\nu
^2}{18}\right)\right]\rrll+{1 \over c^4}\left[v^4 \left(\frac{1349}{3960}
-\frac{4159 \,\nu }{792}+\frac{52409 \,\nu ^2}{1980}-\frac{171539 \,\nu
^3}{3960}\right) +{G^2\,m^2 \over r^2} \left(-\frac{45}{44}-\frac{988
\,\nu}{99}+\frac{9925 \,\nu ^2}{198} \rrrrllll-\frac{8099 \,\nu
^3}{396}\right)+{G\,m \over r} \dot{r}^2 \left(-\frac{23}{396} -\frac{637 \,\nu
}{990}-\frac{1861 \,\nu^2}{990}+\frac{32221 \,\nu ^3}{1980}\right) +{G\,m \over
r} v^2 \left(\frac{1597}{660}-\frac{19381 \,\nu }{990} \rrrrllll+\frac{6307
\,\nu ^2}{198}+\frac{21127 \,\nu^3}{396}\right)\right]\right]
+x_{j}v_{k\ra b}x_{a}{r\,\dot{r} \over c^2}\left[\frac{2}{9}-\frac{10 \,\nu
}{9} +\frac{10 \,\nu ^2}{9}+{1 \over c^2}\left[v^2
\left(\frac{73}{495}-\frac{841 \,\nu }{495} +\frac{3002 \,\nu
^2}{495}\rrrrllll-\frac{3151 \,\nu ^3}{495}\right)+{G\,m \over r}
\left(\frac{133}{66} -\frac{81 \,\nu}{55}-\frac{3914 \,\nu ^2}{165}+\frac{3089
\,\nu ^3}{330}\right)\right] \right]
+v_{jk\ra b}x_{a}{r^2 \over c^2}\left[\frac{7}{45} \left(1-5 \,\nu +5 \,\nu
^2\right) \rrll+{1 \over c^2}\left[v^2 \left(\frac{119}{990}-\frac{259 \,\nu
}{198}+\frac{2219 \,\nu ^2}{495} -\frac{4529 \,\nu ^3}{990}\right)+\dot{r}^2
\left(\frac{14}{165}-\frac{98 \,\nu}{165} +\frac{196 \,\nu ^2}{165}-\frac{98
\,\nu ^3}{165}\right) \rrrlll+{G\,m \over r}\left(\frac{751}{495}-\frac{1792
\,\nu }{495}-\frac{227 \,\nu ^2}{99} +\frac{427 \,\nu^3}{99}\right)\right]
\right] \right\}+ \mathcal{O}\left(\frac{1}{c^5}\right),\\
J_{ijkl}&=-\,\nu\,m\,\Delta\,\epsilon_{ab\la i}\left\{
x_{jkl\ra a}v_{b}\left[1-2 \,\nu+{1 \over c^2}\left[{G\,m \over r}
\left(\frac{15}{11} +\frac{35 \,\nu }{44}-\frac{185 \,\nu ^2}{22}\right)+v^2
\left(\frac{5}{11} -\frac{95 \,\nu }{22}\rrrrllll+\frac{195 \,\nu
^2}{22}\right)\right]\right]
+\frac{5}{22} \,x_{jk}v_{l\ra b}x_{a}{r\,\dot{r} \over c^2} \left(1-4 \,\nu +3
\,\nu ^2\right)
+\frac{4}{11}\,x_{j}v_{kl\ra b}x_{a}{r^2 \over c^2} \left(1-4 \,\nu +3 \,\nu
^2\right) \right\}\nn+ \mathcal{O}\left(\frac{1}{c^4}\right),\\
J_{ijklm}&=\,\nu\,m\,\epsilon_{ab\la i}\left\{
x_{jklm\ra a}v_{b}\left[1-5 \,\nu +5 \,\nu ^2+{1 \over c^2}\left[v^2
\left(\frac{83}{182} -\frac{161 \,\nu }{26}+\frac{317 \,\nu ^2}{13}-\frac{707
\,\nu ^3}{26}\right)\rrrlll+{G\,m \over r} \left(\frac{81}{65}-\frac{138 \,\nu
}{65}-\frac{210\,\nu ^2}{13} +\frac{339 \,\nu ^3}{13}\right)\right]\right]
+\frac{20}{91}\,x_{jkl}v_{m\ra b}x_{a}{r\,\dot{r} \over c^2} \left(1-7 \,\nu
+14 \,\nu ^2-7 \,\nu ^3\right)
\rl+\frac{54}{91}\,x_{jk}v_{lm\ra b}x_{a}{r^2 \over c^2} \left(1-7 \,\nu +14
\,\nu ^2-7 \,\nu ^3\right)\right\}+ \mathcal{O}\left(\frac{1}{c^4}\right),\\
J_{ijklmn}&=-\,\nu\,m\Delta\,\epsilon_{ab\la i}x_{jklmn\ra}x_{a}v_{b}\left(1-4
\,\nu +3 \,\nu ^2\right)+ \mathcal{O}\left(\frac{1}{c^2}\right),\\
J_{ijklmno}&=\,\nu\,m\,\epsilon_{ab\la i}x_{jklmno\ra}x_{a}v_{b} \left(1-7
\,\nu +14 \,\nu ^2-7 \,\nu ^3\right)+ \mathcal{O}\left(\frac{1}{c^2}\right).
\end{align} \end{widetext} \ese
The required gauge moments, the monopolar moment $W$, and two dipolar moments
$W_{i}$ and $Y_{i}$ are finally given by
\bse \label{ellWF:WY} \begin{align} W &=
\frac{1}{3}\,\nu\,m\,r\,\dot{r}+\mathcal{O}\left(\frac{1}{c^2}\right)
\label{ellWF:W0}\,,\\ W_i &= \frac{1}{10}\,\nu\,m\,\Delta\,r^2\left[v^i
-3\,\frac{\dot{r}}{r}\,x^i\right]+ \mathcal{O}\left(\frac{1}{c^2}\right)
\label{ellWF:W1}\,,\\ Y_i &= \frac{1}{5}\,\nu\,m\,\Delta\left[{1
\over2}\frac{G\,m}{r}\,x^i+{1
\over2}\,v^2\,x^i-\frac{3}{2}\,r\,\dot{r}\,v^i\right]+
\mathcal{O}\left(\frac{1}{c^2}\right)\,.  \label{ellWF:Y1} \end{align} \ese
\subsection{The post-Newtonian compact binary dynamics}
\label{ellWF:ellWFbinarydynamics}

Since relations connecting the radiative multipole moment to the source
multipole moment involve time derivatives of the source multipole moments,
computations of various modes will require a knowledge of the equations of
motion (EOM) with the PN accuracy with which one wants to compute various
modes.  Before we write expressions for the EOM, with the PN accuracy required
for the present work, let us recall the definitions of various dynamical
variables as well as some other related results which will be used in
calculations performed here. The binary's relative separation, $x^i$ is given
by
\begin{align} \label{ellWF:x} x^i&=y_1^i-y_2^i=r\,n^i \end{align}       
with $r=|{\bf x}|$, where $\bf x$ is the relative separation vector. Here,
$y_1^i$ and $y_2^i$ are position vectors of the individual components of the
binary and ${n^i}$ are components of the unit vector, $\bf {\hat{n}}$, along
the relative separation vector.\\ 

For the relative velocity and relative acceleration we have

\begin{align} \label{ellWF:v-a} v^i&={{\rm d}x^i \over {\rm
d}t}\,\,\,\,\,\,{\rm and}\,\,\,\,\, \dot{r}={\bf \hat{n}}\cdot {\bf v},\\
a_i&={{\rm d}v^i \over {\rm d}t}={{\rm d^2}x^i \over {\rm d^2}t}.  \end{align}

In addition we also would need expressions for $\dot{v}$ and $\ddot{r}$, which
can be given as

\begin{align} \label{ellWF:vdot-rdot} \dot{v}&={{\bf a}\cdot{\bf v} \over v},\\
\ddot{r}&={1 \over r}\left[\left(v^2-\dot{r}^2\right)+{\bf a}\cdot{\bf
x}\right].  \end{align} with $v=|{\bf v}|$.\\
 
While computing the time derivatives of the source multipole moments, whenever
quantities like $a^i$ or $\dot{v}$ or $\ddot{r}$ appear, they are consistently
replaced by their expressions in terms of variables related to the position and
velocities $(r, \dot{r}, v)$. Computations of various modes at the 3PN order
would require the knowledge of the 3PN EOM governing the compact binary
dynamics. EOM associated with SH coordinates at 3PN order for a system of two
compact objects moving in general orbits is  available in the literature
\cite{BFeom, BDE04, DJSequiv, ABF01, DJSdim} and are given in terms of the
variables related to the position and the velocity of individual constituents
of the binary. Since we are using expressions for the source multipole moments
in the center-of-mass frame of the system, we need the EOM reduced to the CM
frame.  The 3PN accurate expression for relative acceleration, reduced to the
CM frame, in SH coordinates, were obtained in \cite{BI03CM}.  However, as
discussed in the previous section (about using MH or ADM coordinates instead of
SH coordinates), we wish to use EOM reduced to CM frame associated with MH
coordinate which is given in \cite{MW03} and takes the following form,   

\begin{widetext} \begin{align} \label{ellWF:arel} a^i&=-{G\,m \over
r^2}\left\{\left[P_1-{\nu \over c^5}\left(\frac{136}{15} {G^2\,m^2 \over r^2}
\dot{r}+\frac{24}{5} {G\,m \over r} \dot{r} v^2\right)\right]\,n^i
+\left[P_2\,{\dot{r} \over c^2}+{\nu \over c^5}\left(\frac{24}{5}{G^2\,m^2
\over r^2} +\frac{8}{5} {G\,m \over r} v^2\right)\right]\,v^i\right\},
\end{align} \end{widetext} where, \begin{widetext} \bse \label{ellWF:P1-P2}
\begin{align}
P_1&=1+{1 \over c^2}\left[{G\,m \over r} (-4-2 \nu )-\frac{3 \dot{r}^2 \nu }{2}
+v^2 (1+3 \nu )\right]+{1 \over c^4}\left[{G^2\,m^2 \over r^2} \left(9
+\frac{87 \nu }{4}\right) +\dot{r}^4 \left(\frac{15 \nu }{8}-\frac{45 \nu
^2}{8}\right) \rl+v^4 \left(3 \nu -4 \nu ^2\right)+{G\,m \over r}\dot{r}^2
\left(-2-25 \nu -2 \nu ^2\right) +v^2 \left({G\,m \over r} \left(-\frac{13 \nu
}{2}+2 \nu ^2\right) +\dot{r}^2 \left(-\frac{9 \nu }{2}+6
\nu^2\right)\right)\right] \nonumber\\&+{1 \over c^6}\left[{G^3\,m^3 \over r^3}
\left(-16-\frac{1399 \nu }{12} +\frac{41 \pi ^2 \nu }{16}-\frac{71 \nu
^2}{2}\right) +{G\,m \over r} \dot{r}^4 \left(79 \nu -\frac{69 \nu ^2}{2}
-30\nu ^3\right)+\dot{r}^6 \left(-\frac{35 \nu }{16}\rrll+\frac{175 \nu ^2}{16}
-\frac{175 \nu ^3}{16}\right)+{G^2\,m^2 \over r^2} \dot{r}^2
\left(1+\frac{22717 \nu}{168} +\frac{615 \pi ^2 \nu }{64}+\frac{11 \nu ^2}{8}-7
\nu ^3\right)+v^6 \left(\frac{11 \nu }{4} -\frac{49 \nu ^2}{4}\rrll+13 \nu
^3\right)+v^4 \left(\dot{r}^2\left(-\frac{15 \nu }{2} +\frac{237 \nu
^2}{8}-\frac{45 \nu ^3}{2}\right)+{G\,m \over r} \left(\frac{75 \nu }{4} +8 \nu
^2-10 \nu ^3\right)\right)+v^2\left({G^2\,m^2 \over r^2}
\times\rrll\times\left( -\frac{20827 \nu }{840} -\frac{123 \pi ^2 \nu }{64}+\nu
^3\right)+{G\,m \over r} \dot{r}^2 \left(-121 \nu +16 \nu ^2+20
\nu^3\right)+\dot{r}^4 \left(\frac{15 \nu }{2}-\frac{135 \nu ^2}{4}
\rrrlll+\frac{255 \nu ^3}{8}\right)\right)\right],\\
P_2&=-4+2 \nu+{1 \over c^2}\left[v^2 \left(-\frac{15 \nu }{2}-2 \nu
^2\right)+\dot{r}^2 \left(\frac{9 \nu }{2}+3 \nu ^2\right)+{G\,m \over r}
\left(2+\frac{41 \nu }{2} +4 \nu ^2\right)\right]\nonumber\\&+{1 \over
c^4}\left[\dot{r}^4 \left(-\frac{45 \nu }{8}+15 \nu ^2 +\frac{15 \nu
^3}{4}\right)+v^4 \left(-\frac{65 \nu }{8}+19 \nu ^2+6 \nu ^3\right) +{G^2\,m^2
\over r^2}\left(-4-\frac{5849 \nu }{840}-\frac{123 \pi ^2 \nu }{32}\rrll+25 \nu
^2 +8 \nu ^3\right)+{G\,m \over r} \dot{r}^2 \left(\frac{329 \nu }{6}+\frac{59
\nu ^2}{2} +18\nu ^3\right)+v^2 \left(\dot{r}^2 \left(12 \nu -\frac{111 \nu
^2}{4}-12 \nu ^3\right) +{G\,m \over r} \left(-15 \nu \rrrlll-27 \nu ^2-10 \nu
^3\right)\right)\right].  \end{align} \ese \end{widetext}

We now have all the inputs which are needed to compute the instantaneous
expressions for various spherical harmonic modes ($h^{\ell m}$) associated with
3PN gravitational waveforms of GW signals from CCBs moving in general orbits.
With this motivation, we shall proceed towards the next section where we shall
present our results.


\section{Instantaneous terms in the 3PN gravitational waveform for CCBs in 
general orbits}
\label{ellWF:ellWF3pnsphhar}

Combing Eq.~\eqref{ellWF:UV} and Eq.~\eqref{ellWF:hlm-modesep} we can write the
instantaneous part of various modes as

\begin{align}\label{ellWF:hlm-modesep-inst} h^{\ell m}_{\rm inst}
&=-\frac{G}{{\sqrt 2}\,R\,c^{\ell+2}}\frac{4}{\ell!}
\,\sqrt{\frac{(\ell+1)(\ell+2)}{2\ell(\ell-1)}}\,\alpha_L^{\ell m}\,U_L^{\rm
inst} \nn \quad\quad\quad\quad {\rm if}\,\ell+m\,\,{\rm is\, even},\\ h^{\ell
m}_{\rm inst} &=-\frac{i\,G}{{\sqrt 2}\,R\,c^{\ell+3}}\frac{8}{\ell!}
\,\sqrt{\frac{\ell(\ell+2)}{2(\ell+1)(\ell-1)}}\,\alpha_L^{\ell m}\,V_L^{\rm
inst} \nn \quad\quad\quad\quad {\rm if}\,\ell+m\,\,{\rm is\, odd}.  \end{align}

Relations connecting the instantaneous part of STF radiative moments ($U_L^{\rm
inst}$ and $V_L^{\rm inst}$) to the source multipole moments have been listed
in the previous section.  This allows one to write the instantaneous part of
various modes ($h^{\ell m}_{\rm inst}$) in terms of the source multipole
moments. With expressions for the source multipole moments for CCBs moving in
general orbits and their relevant time derivatives, one can write expressions
for various modes in terms of dynamical variables related to the position and
velocity ($r$, $\dot{r}$, $\phi$, $v$).\footnote{Alternatively, one can also
compute various modes associated with the gravitational waveform using
polarization waveforms (see Sec. II and IX of Ref.~\cite{BFIS08} for the
details).}

Again, since $v^2=\dot{r}^2+r^2\dot{\phi}^2$,  we can write various modes of
the waveform in terms of the dynamical variables, namely, the radial separation
$(r)$, radial velocity $(\dot{r})$, orbital phase $(\phi)$, and the angular
velocity $(\dot{\phi})$ . The structure of $h^{\ell m}$ reads

\be \label{ellWF:hlm} h^{\ell m}_{\rm inst}={4\,G\,m\,\nu \over c^4 R}{\sqrt
{\pi \over 5}}\,e^{-i\,m\,\phi}\,\hat{H}^{\ell m}_{\rm inst}.  \ee

For the dominant mode ($\ell=2, m=2$), with 3PN accuracy, various PN pieces of
the coefficient $\hat{H}^{22}_{\rm inst}$ read 
\begin{widetext} \bse \label{ellWF:H22} \begin{align} (\hat{H}^{22}_{\rm
inst})_{\rm Newt}&=\frac{G m}{r}+r^2 \dot{\phi }^2+2 i r \dot{r} \dot{\phi
}-\dot{r}^2,\\
(\hat{H}^{22}_{\rm inst})_{\rm 1PN}&={1 \over c^2}\left[\frac{G^2
m^2}{r^2}\left(-5+\frac{\,\nu }{2}\right) +\frac{G m
\dot{r}^2}{r}\left(-\frac{15}{14}-\frac{16 \,\nu
}{7}\right)+\left(-\frac{9}{14} +\frac{27 \,\nu}{14}\right)
\dot{r}^4+r\left(\frac{9 i}{7}\rrll-\frac{27 i \,\nu }{7}\right) \dot{r}^3
\dot{\phi } +G m r \left(\frac{11}{42}+\frac{26 \,\nu }{7}\right) \dot{\phi
}^2+r^4 \left(\frac{9}{14}-\frac{27 \,\nu }{14}\right) \dot{\phi }^4 +\dot{r}
\left(G m \left(\frac{25 i}{21}+\frac{45 i \,\nu }{7}\right) \dot{\phi }
\rrll+r^3\left(\frac{9 i}{7}-\frac{27 i \,\nu }{7}\right) \dot{\phi
}^3\right)\right],\\
(\hat{H}^{22}_{\rm inst})_{\rm 2PN}&={1 \over c^4}\left[\frac{G^3 m^3}{r^3}
\left(\frac{757}{63}+\frac{181 \,\nu }{36} +\frac{79 \,\nu
^2}{126}\right)+\left(-\frac{83}{168}+\frac{589 \,\nu }{168}-\frac{1111 \,\nu
^2}{168}\right) \dot{r}^6+r \left(\frac{83 i}{84} \rrll-\frac{589 i \,\nu
}{84}+\frac{1111 i \,\nu ^2}{84}\right) \dot{r}^5 \dot{\phi} +G^2 m^2
\left(-\frac{11891}{1512}-\frac{5225 \,\nu }{216}+\frac{13133 \,\nu
^2}{1512}\right) \dot{\phi }^2+G m r^3 \left(\frac{835}{252}\rrll+\frac{19
\,\nu }{252}-\frac{2995 \,\nu ^2}{252}\right) \dot{\phi }^4+r^6
\left(\frac{83}{168}-\frac{589 \,\nu }{168} +\frac{1111 \,\nu ^2}{168}\right)
\dot{\phi }^6 +\dot{r}^4 \left(\frac{G m}{r}\left(-\frac{557}{168}+\frac{83
\,\nu }{21} \rrrlll+\frac{214 \,\nu ^2}{21}\right)+r^2
\left(-\frac{83}{168}+\frac{589 \,\nu }{168} -\frac{1111 \,\nu
^2}{168}\right)\dot{\phi }^2\right)+\dot{r}^3 \left(G m \left( \frac{863
i}{126}-\frac{731 i \,\nu }{63} \rrrlll-\frac{211 i \,\nu ^2}{9}\right)
\dot{\phi } +r^3 \left(\frac{83 i}{42}-\frac{589 i \,\nu }{42}+\frac{1111 i
\,\nu ^2}{42}\right) \dot{\phi }^3\right)+\dot{r}^2\left(\frac{G^2
m^2}{r^2}\left(\frac{619}{252} -\frac{2789 \,\nu}{252}\rrrlll-\frac{467 \,\nu
^2}{126}\right) +G m r \left(\frac{11}{28} -\frac{169 \,\nu }{14}-\frac{58
\,\nu ^2}{21}\right) \dot{\phi }^2 +r^4\left(\frac{83}{168}-\frac{589 \,\nu
}{168} +\frac{1111 \,\nu ^2}{168}\right) \dot{\phi }^4\right) \rl+\dot{r}
\left(\frac{G^2 m^2}{r}\left(-\frac{773 i}{189}-\frac{3767 i \,\nu
}{189}+\frac{2852 i \,\nu ^2}{189}\right)\dot{\phi }+G m r^2 \left(\frac{433
i}{84} +\frac{103 i \,\nu }{12} -\frac{1703 i \,\nu ^2}{84}\right) \dot{\phi
}^3\rrll+r^5 \left(\frac{83 i}{84}-\frac{589 i \,\nu }{84}+\frac{1111 i \,\nu
^2}{84} \right)\dot{\phi }^5\right)\right],\\
(\hat{H}^{22}_{\rm inst})_{\rm 2.5PN}&={1 \over c^5}\left[-\frac{122 G^2 m^2
\,\nu  \dot{r}^3}{35 r^2} -\frac{468 i G^3 m^3 \,\nu  \dot{\phi }}{35
r^2}+\frac{184 i G^2 m^2 \,\nu  \dot{r}^2 \dot{\phi }}{35 r}-\frac{316}{35} i
G^2 m^2 r \,\nu  \dot{\phi }^3 \rl+\dot{r} \left(\frac{2 G^3 m^3 \,\nu }{105
r^3}-\frac{121}{5} G^2 m^2 \,\nu  \dot{\phi }^2\right)\right],\\
(\hat{H}^{22}_{\rm inst})_{\rm 3PN}&={1 \over c^6}\left[\frac{G^4
m^4}{r^4}\left(-\frac{512714}{51975}+\left( -\frac{1375951}{13860}+\frac{41 \pi
^2}{16}\right) \,\nu +\frac{1615 \,\nu ^2}{616} +\frac{2963
\,\nu^3}{4158}-\frac{214}{105} \log \left[\frac{r}{r_0}\right]\right)
\rl+\left(-\frac{507}{1232}+\frac{6101 \,\nu }{1232} -\frac{12525
\,\nu^2}{616}+\frac{34525 \,\nu ^3}{1232}\right) \dot{r}^8+r \left( \frac{507
i}{616}-\frac{6101 i \,\nu }{616}+\frac{12525 i \,\nu ^2}{308}
\rrll-\frac{34525 i \,\nu^3}{616}\right) \dot{r}^7 \dot{\phi }+\frac{G^3
m^3}{r} \left(\frac{42188851}{415800}+\left(\frac{190703}{3465} -\frac{123 \pi
^2}{64}\right) \,\nu -\frac{18415 \,\nu^2}{308}+\frac{281473 \,\nu^3}{16632}
\rrll-\frac{214}{15} \log\left[\frac{r}{r_0}\right]\right) \dot{\phi }^2+G^2
m^2 r^2 \left(\frac{328813}{55440}-\frac{374651 \,\nu }{33264}+\frac{249035
\,\nu ^2}{4158} -\frac{1340869 \,\nu ^3}{33264}\right) \dot{\phi }^4\rl+G m r^5
\left(\frac{12203}{2772}-\frac{36427 \,\nu }{2772}-\frac{13667 \,\nu ^2}{1386}
+\frac{49729 \,\nu ^3}{924}\right) \dot{\phi }^6+r^8
\left(\frac{507}{1232}-\frac{6101 \,\nu }{1232}+\frac{12525 \,\nu ^2}{616}
\right.\right.\nonumber \\ & \left.\left.  -\frac{34525 \,\nu ^3}{1232}\right)
\dot{\phi }^8+\dot{r}^4 \left(\frac{G^2 m^2 }{r^2}
\left(-\frac{92567}{13860}+\frac{7751 \,\nu}{396}+\frac{400943 \,\nu ^2}{11088}
+\frac{120695 \,\nu ^3}{3696}\right) \rrll+G m r
\left(-\frac{42811}{11088}+\frac{6749 \,\nu }{1386}+\frac{19321 \,\nu^2}{693}
-\frac{58855 \,\nu ^3}{1386}\right) \dot{\phi }^2\right)+\dot{r}^6
\left(\frac{G m}{r} \left(-\frac{5581}{1232}+\frac{4694 \,\nu
}{231}\rrrlll-\frac{3365 \,\nu^2}{462}-\frac{1850 \,\nu ^3}{33}\right)+r^2
\left(-\frac{507}{616}+\frac{6101 \,\nu }{616} -\frac{12525 \,\nu ^2}{308}
+\frac{34525 \,\nu ^3}{616}\right)\dot{\phi}^2\right)\rl +\dot{r}^5\left(G m
\left(\frac{17233 i}{1848} -\frac{31532 i \,\nu }{693}+\frac{65575 i \,\nu
^2}{2772}+\frac{85145 i \,\nu ^3}{693}\right) \dot{\phi} +r^3 \left(\frac{1521
i}{616}-\frac{18303 i \,\nu }{616}\rrrlll+\frac{37575 i \,\nu ^2}{308}
-\frac{103575 i \,\nu ^3}{616}\right) \dot{\phi }^3\right)+\dot{r}^3
\left(\frac{G^2 m^2}{r}\left(\frac{39052 i}{3465}-\frac{154114 i \,\nu }{2079}
-\frac{246065 i \,\nu ^2}{4158}\rrrlll-\frac{365725 i \,\nu ^3}{4158}\right)
\dot{\phi }+G m r^2 \left(\frac{13867 i}{792}-\frac{191995 i \,\nu }{2772}
-\frac{8741 i \,\nu ^2}{5544}+\frac{52700 i \,\nu ^3}{231}\right) \dot{\phi }^3
+r^5 \left(\frac{1521 i}{616}\rrrlll-\frac{18303 i \,\nu }{616}+\frac{37575 i
\,\nu ^2}{308} -\frac{103575 i \,\nu ^3}{616}\right) \dot{\phi
}^5\right)+\dot{r}^2 \left(\frac{G^3 m^3 }{r^3}\left(\frac{913799}{29700}
+\left(\frac{174679}{2310}+\frac{123 \pi ^2}{32}\right) \,\nu
\rrrlll-\frac{158215 \,\nu ^2}{2772} -\frac{12731 \,\nu
^3}{4158}-\frac{428}{105} \log\left[\frac{r}{r_0}\right]\right)+G^2 m^2
\left(\frac{20191}{18480} -\frac{3879065 \,\nu }{33264}-\frac{411899 \,\nu
^2}{8316} \right.\right.\right.\nonumber \\ & \left.\left.\left.  -\frac{522547
\,\nu ^3}{33264}\right) \dot{\phi }^2+G m r^3 \left(\frac{381}{77}-\frac{101237
\,\nu }{2772} +\frac{247505 \,\nu ^2}{5544}+\frac{394771 \,\nu ^3}{5544}\right)
\dot{\phi }^4+r^6 \left(\frac{507}{616}\rrrlll-\frac{6101 \,\nu
}{616}+\frac{12525 \,\nu ^2}{308} -\frac{34525 \,\nu ^3}{616}\right) \dot{\phi
}^6\right)+\dot{r} \left(\frac{G^3 m^3 }{r^2}\left(-\frac{68735
i}{378}+\left(-\frac{57788 i}{315} +\frac{123 i \pi ^2}{32}\right) \,\nu
\rrrlll-\frac{701 i \,\nu ^2}{27}+\frac{11365 i \,\nu ^3}{378} +\frac{428}{21}
i \log\left[\frac{r}{r_0}\right]\right) \dot{\phi } +G^2 m^2 r
\left(\frac{91229 i}{13860}+\frac{97861 i \,\nu }{4158}+\frac{919811 i \,\nu
^2}{8316}\rrrlll-\frac{556601 i \,\nu ^3}{8316}\right) \dot{\phi }^3 +G m r^4
\left(\frac{6299 i}{792}-\frac{68279 i \,\nu }{5544}-\frac{147673 i
\,\nu^2}{2772} +\frac{541693 i \,\nu ^3}{5544}\right) \dot{\phi }^5+r^7
\left(\frac{507 i}{616} \rrrlll-\frac{6101 i \,\nu }{616}+\frac{12525 i \,\nu
^2}{308}-\frac{34525 i \,\nu ^3}{616}\right) \dot{\phi }^7\right)\right].
\end{align} \ese \end{widetext}
As mentioned in Sec.~\ref{ellWF:ellWFintro}, since expressions for various
modes are very large and would run over several pages, we have decided to
provide an additional file (Hlm-GenOrb.m) containing expressions for all the
modes which will be made available along with the paper. Finally,
circular-orbit limit of the instantaneous $h_{\ell m}$ can be obtained by
replacing related expressions for $\dot{\phi}(=\omega)$, $\dot{r}$ and $r$
given in Sec.~IV of \cite{BFIS08}. \\

Note that $h_{\ell m}$ can directly be used to write the polarization waveforms
($h_+$, $h_{\times}$) using the standard decomposition of $h_{+}$ and
$h_{\times}$ in terms of spherical harmonic modes of spin weight -2 given in
Ref.~\cite{BFIS08, K08},
\begin{equation} \label{ellWF:spinw} h_+ - i h_\times =
\sum^{+\infty}_{\ell=2}\sum^{\ell}_{m=-\ell} h^{\ell m}\,Y^{\ell
m}_{-2}(\Theta,\Phi)\,, \end{equation}
where $Y^{\ell m}_{-2}$'s (the spin-weighted spherical harmonics of weight
$-2$) are functions of the spherical angles $(\Theta,\Phi)$ defining the
binary's location and given as
\begin{widetext}\begin{subequations}\label{ellWF:harm}\begin{align} Y^{\ell
m}_{-2} &= \sqrt{\frac{2\ell+1}{4\pi}}\,d^{\,\ell m}_{\,2}(\Theta)\,e^{i \,m
\,\Phi},\\d^{\,\ell m}_{\,2} &= \sum_{k=k_1}^{k_2}\frac{(-)^k}{k!}
\frac{\sqrt{(\ell+m)!(\ell-m)!(\ell+2)!(\ell-2)!}}
{(k-m+2)!(\ell+m-k)!(\ell-k-2)!}\left(\cos\frac{\Theta}{2}\right)^{2\ell+m-2k-2}
\!\!\!\left(\sin\frac{\Theta}{2}\right)^{2k-m+2}\,,
\end{align}\end{subequations}\end{widetext}
with $k_1=\mathrm{max}(0,m-2)$ and $k_2=\mathrm{min}(\ell+m,\ell-2)$.\\ 

Further, the polarization waveform can be used to write the
transverse-traceless part of the radiation field ($h_{ij}$) by using the
following relation~\cite{BFIS08, K08}
\begin{equation} \label{ellWF:hij} h_{ij}^{\rm TT}=2
\left(h_+\,e_{ij}^{+}+h_{\times}\,e_{ij}^{\times}\right)\,, \end{equation}
where \bse\label{ellWF:epec} \begin{align} \label{ellWF:ep} e_{ij}^{+}&={1
\over2}\left(P_i P_j-Q_i Q_j\right)\,,\\ \label{ellWF:ep} e_{ij}^{\times}&={1
\over2}\left(P_i Q_j+P_j Q_i\right)\,. \end{align} \ese
Here {\bf P} and {\bf Q} are the two unit polarization vectors and they have
been chosen following the convention used in \cite{BFIS08}. 

\section{Quasi-Keplerian representation}
\label{ellWF:ellWFQKR}

In the previous section, we presented general orbit expressions for the
dominant mode of the gravitational waveform from a compact binary system. In
this section we aim to specialize to the case of compact binary systems in
quasi-elliptical orbits. Expressions for various $h_{\ell m}$ describing the
radiation from binaries in quasi-elliptical orbits can simply be obtained by
using relations connecting generic dynamical variable $r$, $\dot{r}$, $\phi$
and $\dot{\phi}$ to a set of parameters associated with elliptical orbits in
general orbit expressions for various modes.  Such relations can be established
by using the generalized quasi-Keplerian (QK) representation of the
conservative dynamics of the binary moving in eccentric orbits, which indeed is
available to us due to the work of Memmesheimer, Gopakumar and Sch{\"a}fer
(hereafter MGS) \cite{MGS04}.\\

The QK representation was first introduced by Damour and Deruelle \cite{DD85}
and dealt with the binary dynamics at 1PN order. The generalized QK
representation at 2PN order in ADM-type coordinates was given in
Refs.~\cite{DS88, SW93, Wex95}. MGS provides the 3PN generalized QK
representation in both the ADM and MH coordinates, which involve expressions of
the orbital elements associated with the orbit of the binary in terms of the
conserved energy and orbital angular momentum of the binary. Before we get into
the details of the parametrization, we first summarize equations describing the
radial and angular motion of the binary in terms of various orbital elements
associated with elliptical orbits (see Refs.~\cite{ABIQ07, KG06} for details).
In the parametric form, the radial separation, $r$, is given by
\be \label{ellWF:r} r=a_r\left(1-e_r\,\cos u\right), \ee
where $a_r$ is the semi-major axis of the orbit and $e_r$ is the eccentricity
of the orbit (both labeled after the radial coordinate, $r$). The quantity $u$
is called eccentric anomaly and at the 3PN order it is related to the mean
anomaly ($l$) by the relation
\be \label{ellWF:l-u} l=u-e_t\,\sin u+f_t\,\sin
V+g_{t}\left[V-u\right]+i_t\,\sin 2 V +h_t\,\sin 3 V.  \ee
The orbital phase, $\phi$, at the 3PN order reads
\begin{align} \label{ellWF:Phi} \phi&=\phi_P+K\,\left[V+f_\phi\,\sin 2
V+g_\phi\,\sin 3 V +i_\phi\,\sin 4 V \rl + h_\phi\,\sin 5 V\right], \end{align}
where $\phi_P$ is the initial phase at the first passage of the periastron and
$V$ is the true anomaly that takes the form 
\be \label{ellWF:Vuephi} V\equiv V(u,e_\phi)=2\,\arctan\left[{\sqrt{1+e_{\phi}
\over 1-e_{\phi}}\,\tan \left({u \over2}\right)}\right].  \ee    
Also, the mean anomaly, $l$, is related to the time as
\be \label{ellWF:l-n} l=n\left(t-t_P\right), \ee
where $t_P$ is the instant of the first passage at the periastron and
$n=2\,\pi/P$ is the mean motion with P being the orbital period.\\
In addition to this, expressions for the radial and angular velocity can be
given as
\bse \label{ellWF:rdot-phidot} \begin{align} \label{ellWF:rdot}
\dot{r}&=a_r\,e_r\,\sin u\,\left({\partial l \over\partial
u}\right)^{-1}\,{\partial l \over\partial t},\\
\label{ellWF:Phidot} \dot{\phi}&=K\left(1+2\,f_{\phi}\,\cos 2
V+3\,g_{\phi}\,\cos 3 V+4\,i_{\phi}\,\cos 4 V \rl+5\,h_{\phi}\,\cos 5
V\right){\partial V \over\partial u} \left({\partial l \over\partial
u}\right)^{-1}{\partial l \over\partial t}. \end{align} \ese It may be seen
from Eqs.~\eqref{ellWF:l-u}, \eqref{ellWF:Vuephi} and \eqref{ellWF:l-n} that
\bse \label{ellWF:dldt-dldu} \begin{align} \label{ellWF:dldt} {\p l \over\p
t}&=n,\\ \label{ellWF:dldu} {\p l \over\p u}&=1-e_t\,\cos u+f_t\,\cos V\,{\p V
\over\p u}+g_t({\p V \over\p u}-1) \nn+2\,i_t\,{\cos 2 V}\,{\p V \over\p
u}+3\,h_t\,{\cos 3 V}\,{\p V \over\p u}.\\
\label{ellWF:dVdu} {\p V \over\p u}&={\left(1-{e_\phi}^2\right)^{1/2}
\over(1-e_\phi\,\cos u)}.  \end{align} \ese

In the above, $e_\phi$ and $e_t$ denote eccentricities related to the
coordinates $\phi$ and $t$, respectively. $K$ is related to the advance of the
periastron per orbit and is given by $K=\Phi/(2\,\pi)$, where $\Phi$ is the
angle of return to the periastron. In this parametrization, $f_t$, $f_\phi$,
$g_t$ and $g_\phi$ contribute both at 2PN and 3PN order whereas $i_t$,
$i_\phi$, $h_t$ and $h_\phi$ contribute only at the 3PN order (see
Ref.~\cite{ABIQ07} for related details).\\

Once we have written equations connecting the generic dynamical variables ($r$,
$\dot{r}$, $\phi$, and $\dot{\phi}$) to the orbital elements of the elliptical
orbit, we can use inputs from MGS to express them in terms of a suitable set of
parameters of our choice. The main result of MGS is that, it provides 3PN
accurate expressions for various orbital elements ($a_r, e_t, e_r, e_\phi...$)
associated with the elliptical orbits in terms of the corresponding conserved
energy per unit reduced mass ($E$) and the parameter $h$, related to the
reduced angular momentum ($J$), by $h=J/G\,m$.\footnote{In Ref.~\cite{ABIQ07},
which uses results obtained in MGS, orbital elements are expressed in terms of
the parameters $\left\{\epsilon, j\right\}$ instead of $\left\{E, h\right\}$,
where $\epsilon$ and $j$ are defined as $\epsilon=-2\,E/c^2$ and
$j=-2\,E\,h^2$.} Using these relations one can express the dynamical variables
$r$, $\dot{r}$, $\phi$ and $\dot{\phi}$ in terms of $E$, $h$ and $u$. Here one
should note that this is not the only way in which the orbital dynamics can be
parametrized. In fact, one can re-express $E$ and $h$ in terms of any of the
two orbital elements to write equations describing the orbital motion of the
binary; however a parametrization involving gauge invariant parameters is
sometimes preferred as such parametrization is suitable for making comparisons
with related numerical results. This led MGS to use a parametrization involving
$n$ and $K=\Phi/(2\,\pi)$ (both are independent of the coordinate system used
when expressed in terms of $E$ and $h$ in order to describe the orbital motion
of the binary in elliptical orbits).\footnote{In fact, MGS uses $x_{\rm MGS}=
\left({G\,m\,n/c^3}\right)^{2/3}$ and the parameter $k'=(K-1)/3$.} Here, $n$ is
the mean motion and $K=\Phi/(2\,\pi)$ denotes the angle of the advance of the
periastron per orbital revolution. In a work related to the phasing of the GWs
from inspiralling compact binary in elliptical orbit due to Damour, Gopakumar
and Iyer \cite{DGI04}, the orbital dynamics has been described using $n$ and
$e_t$ as parametrizing variables. Following the conventions of \cite{DGI04},
K{\"o}nigsd{\"o}fer and Gopakumar \cite{KG06} provided the 3PN accurate
expressions for $r$, $\dot{r}$, $\phi$ and $\dot{\phi}$ in terms of $n$, $e_t$
and $u$.\footnote{In fact, \cite{KG06} is the extension of \cite{DGI04} and
discusses the 3PN conservative dynamics of the binaries in elliptical orbits.}
Reference~\cite{ABIQ07} makes an alternative choice of parametrization in terms
of variables $x$ and $e_t$, where $x$ is related to the orbital frequency
$\omega$ by, $x=(G\,m\,\omega/c^3)^{2/3}$, and is independent of the choice of
the coordinate system used.\footnote{Parameters $n$ and $\omega$ are related by
$n=K\omega$, and $K$ has been defined above.} The choice $\left\{x,
e_t\right\}$ as parametrizing variables leads to expressions which can be
reduced to those related to the circular orbit case ($e_t\rightarrow0$) which
uses $x$ as the expansion parameter. In addition to this, in another related
work Hinder {\it et al.}  \cite{2010PhRvD..82b4033H} compared the two
parameterizations and concluded that the choice of $x$ as compared to $n$
provides better agreement with NR results. They separately discuss these two PN
models (based on the choice of parametrization as $x$ or $n$) and call them
x-model and n-model. Note that although \cite{2010PhRvD..82b4033H} uses 3PN QK
representation describing the conserved dynamics it used only 2PN accurate
dissipative dynamics and, hence, when used to obtain quasi-circular orbit limit
of the orbital phase it shows some deviations from related standard results as
was pointed out in Ref.~\cite{Huerta:2014eca}. As discussed below our results
must be coupled with 3PN evolution equations for QK variables given in
Ref.~\cite{Arun:2009mc} and, hence, would be more suitable for obtaining
quasi-circular limits of the amplitude and phase of various modes.\\

Following the arguments presented above we choose to parametrize the dynamical
variables $\{r, \dot{r}, \phi, \dot{\phi}\}$ in terms of the QK variables $\{x,
e_t, u\}$ which can further be used to write expressions for various $h_{\ell
m}$ in terms of the QK variables. Note that Ref.~\cite{2010PhRvD..82b4033H}
already lists the 3PN expressions for $r$ and $\dot{\phi}$ in terms of $\{x,
e_t, u\}$; however, they numerically differentiate $r$ to obtain $\dot{r}$ and
perform numerical integration of $\dot{\phi}$ to obtain $\phi$ in order to
avoid the use of long and complicated expressions for ${\dot r}$ and $\phi$ in
their numerical code. As mentioned above, Ref.~\cite{KG06} lists 3PN accurate
expressions for $\{r, \dot{r}, \phi, \dot{\phi}\}$ in terms of the QK variables
$\{n, e_t, u\}$ in MH coordinates (see Eqs.~(23)-(27) there). In fact, they use
a variable $\xi$ related to $n$ by $\xi=(G m n/c^3)$. Hence to obtain related
expressions parametrized in terms of $\{x, e_t, u\}$, all we need to know is how
$\xi$ is related to $x$ and $e_t$.  The relation between $\xi$ and our QK
variables $x$ and $e_t$ with 3PN accuracy in MH coordinates is given as 
\begin{widetext} \begin{align} \label{ellWF:xi2x} \xi &= {x^{3/2}
\over(1-e_t^2)^3}\Biggl\{1-3 e_t^2+3 e_t^4-e_t^6+x \left(-3+6 e_t^2-3
e_t^4\right)+x^2 \left[-\frac{9}{2}+7 \nu +\left(-\frac{33}{4}-\frac{\nu
}{2}\right) e_t^2+\left(\frac{51}{4}-\frac{13 \nu }{2}\right)
e_t^4\right]\nn+x^3 \left[\frac{3}{2}+\nu  \left(\frac{457}{4}-\frac{123 \pi
^2}{32}\right)-7 \nu ^2+\left(-\frac{267}{4}+\nu  \left(\frac{279}{2}-\frac{123
\pi ^2}{128}\right)-40 \nu ^2\right) e_t^2+\left(-\frac{39}{2}+\frac{55 \nu
}{4}-\frac{65 \nu ^2}{8}\right) e_t^4\rl+\sqrt{1-e_t^2} \left(-15+6 \nu
+(-30+12 \nu ) e_t^2\right)\right] \Biggr\}\,.     \end{align} \end{widetext}

Using the above in Eqs.~(23)-(27) of Ref.~\cite{KG06} one can write expressions
for  $r$, $\dot{r}$, $\phi$ and $\dot{\phi}$ in terms of the variables $\{x,
e_t, u\}$ and then use them to obtain the expressions for the spherical
harmonic modes of the waveform for quasi-elliptical orbits.\\ 

We now are in a position to use the expressions for $r$, $\dot{r}$, $\phi$ and
$\dot{\phi}$ in terms of the variables $\{x, e_t, u\}$ to reexpress the
instantaneous part of spin-weighted spherical harmonic modes ($h^{\ell m}_{\rm
inst}$) in terms of parameters $x$, $e_t$ and $u$.  Just like the expressions
for various $h_{\ell m}$ in general orbits we find that the expressions for
various modes in QK representation is too large to be given in the main text of
the paper. Hence we will just provide the expression for the dominant mode here
and list all the relevant modes in an additional file (Hlm-EllOrb.m). In
addition, since even the $h_{22}$ expression runs over many pages we provide
the PN structure of $h_{22}$ in the main text of the paper and list explicit
expressions for various PN pieces in Appendix A to maintain the flow of
discussion in the main text of the paper.\\

The structure of various modes, $h^{\ell m}$, remains the same as in
Eq.\eqref{ellWF:hlm}; however, now the coefficients $\hat{H}^{\ell m}_{\rm
inst}$ and $\phi$ are functions of the parameters $\{x, e_t, u\}$, 
\be \label{ellWF:hlm-QKR} h^{\ell m}_{\rm inst}={4\,G\,m\,\nu\,x \over c^2
R}{\sqrt {\pi \over 5}}\,e^{-i\,m\,\phi}\,\hat{H}^{\ell m}_{\rm inst}.  \ee
The instantaneous part of the dominant mode ($h^{22}_{\rm inst}$) reads 
\begin{align} \label{ellWF:h22-QKR} h^{22}_{\rm inst}&={4\,G\,m\,\nu\,x \over
c^2 R}{\sqrt {\pi \over 5}}\,e^{-2\,i\,\phi}\,\hat{H}^{22}_{\rm inst}\,,
\end{align}
with 
\begin{align} \label{ellWF:H22-QKR-decomp} H^{22}_{\rm inst}&= H^{22}_{\rm
Newt}+H^{22}_{\rm 1PN} + H^{22}_{\rm 2PN}+H^{22}_{\rm 2.5PN} + H^{22}_{\rm
3PN}\,, \end{align} 
where various PN pieces appearing in Eq.~\eqref{ellWF:H22-QKR-decomp} are listed in
Appendix~A. Before we move to the concluding section, we would like to make a
few remarks about the results presented here.\\ 

We observe logarithmic dependences on the arbitrary length scale $r_0$ in
Eq.~\eqref{ellWF:H22-QKR-3PN} through the quantity $x_0$ which is related to
$r_0$ by $x_0=(G m/c^2 r_0)$. This dependence is due to the presence of terms
involving logarithms of $r_0$ in the expression for the mass quadrupole moment
($I_{ij}$) given by Eqs.~\eqref{ellWF:I2}-\eqref{ellWF:A1-A3}.  As was discussed
in Sec.~\ref{ellWF:sourcemoms} we would expect such dependences to disappear
from final expression for various modes (for instance see Ref.~\cite{BFIS08}
which lists $h_{\ell m}$ for binaries in quasi-circular orbits). As has been
observed in Refs.~\cite{B98tail, BFIS08, K08, ABIQ07}, it turns out the
hereditary contribution has equal and opposite dependences on the arbitrary
length scale $r_0$ and cancels out from the final expression. Since we do not
provide hereditary contributions in this paper, such cancellation can not be
shown here. However, we explicitly show this cancellation in \cite{Mishra14}
which deals with computation of hereditary effects in various modes at 3PN
order.\\

In this paper we used the 3PN accurate QK representation which describes
the conserved dynamics of CCBs in eccentric orbits to obtain various modes in terms
of QK variables ($x$, $e_t$). However, it should be noted that the parameters
$x$, $e_t$ evolve with time over radiation reaction time scales and these
secular effects start to show at 2.5PN order \cite{KG06, Arun:2009mc}. Hence,
in order to correctly account for the reactive dynamics of the binary at 3PN
order our results should always be coupled with equations describing secular
time-evolution of $x$ and $e_t$. Equations describing the secular evolution of
orbital elements with relative 3PN accuracies were presented in
Ref.~\cite{Arun:2009mc} (see Sec. VI there). Evolution equations (due to
instantaneous terms in energy and angular momentum loss) for orbital frequency
($d\omega/dt$ (related to $x$ by $\omega=(c^3\,x^{3/2}/G m)$) and for
time-eccentricity parameter ($de_t/dt$) in terms of $x$ and $e_t$ have been
listed as Eqs.~(6.14)-(6.15) and Eqs.~(6.18)-(6.19), respectively, in
Ref.~\cite{Arun:2009mc}.\\

\section{Summary and concluding remarks}
\label{ellWF:ellWFconclusion}

In this paper we presented computations of the instantaneous contributions to
all the relevant modes of the 3PN accurate gravitational waveform of the GW
signal from nonspinning coalescing compact binaries in general orbits. The
expression for the instantaneous part of the dominant mode ($h^{\ell m}_{\rm
inst}$), in terms of the variables $r$, $\dot{r}$, $\phi$ and $\dot{\phi}$, has
been given by Eqs.~\eqref{ellWF:hlm}-\eqref{ellWF:H22} above, whereas
expressions for other subdominant modes (along with the dominant mode) have
been listed in a separate file (Hlm-GenOrb.m) that is being made available
along with the paper.  Next, we specialized to the case of CCBs in
quasi-elliptical orbits using the 3PN quasi-Keplerian representation of the
conserved dynamics of compact binaries in eccentric orbits in
Sec.~\ref{ellWF:ellWFQKR}. Related 3PN accurate expressions for the
instantaneous part of the dominant mode ($h^{\ell m}_{\rm inst}$), in terms of
the variables, namely, the time-eccentricity $e_t$, a PN parameter $x$ and
eccentric anomaly $u$, is given by
Eqs.~\eqref{ellWF:h22-QKR}-\eqref{ellWF:H22-QKR-decomp} and
Eq.~\eqref{ellWF:H22-QKR}. The expressions for other sub-dominant modes (along
with the dominant mode) have been listed in a separate file (Hlm-EllOrb.m) that
is being made available along with the paper.\\ 

We once again remind the readers that the results presented here only account
for the contributions from the instantaneous terms in the waveform which must
be complemented by computations accounting for the hereditary effects. Our
investigations suggest that it is not possible to provide closed-form
analytical expressions for the hereditary terms for binaries moving in general
orbits. Moreover, even for the special case of CCBs in quasi-elliptical orbits
it may not be possible to have closed-form analytical expressions for the
hereditary terms valid for systems with arbitrary eccentricities. However, we
find that such computations can be performed assuming an expansion in the
eccentricity parameter ($e_t$) \cite{Mishra14}.  Unlike the results presented
in this paper which can be applied to a binary with arbitrary eccentricity,
results of \cite{Mishra14} can only be applied to systems with small
eccentricities.  However, the positive side of the work is that we shall have
complete 3PN analytical expression for the waveform for binaries in
quasi-elliptical orbits that can be used for comparison with related numerical
relativity results, which is one of the motivations for high PN order
computations of the gravitational waveforms.  In addition, with complete
waveforms at hand one would be able to write complete polarization waveforms
(as discussed in Sec.~\ref{ellWF:ellWF3pnsphhar}) which would be useful for
data-analysis purposes.\\

\begin{acknowledgments}

We thank L Blanchet and G Faye for useful discussions. CKM thanks Chennai
Mathematical Institute for hospitality. KGA thanks Raman Research Institute for
hospitality. KGA's research was partly funded by a grant from Infosys
Foundation. All the calculations reported in this paper were carried out with
the aid of the algebraic computing software {\tt MATHEMATICA}.

\end{acknowledgments}

\appendix
\section{Various PN pieces of the coefficient $\hat{H}^{22}$ associated with the dominant mode $h^{22}$}
\label{ellWF:apndx1}
\begin{widetext} \bse \label{ellWF:H22-QKR} \begin{align} (\hat{H}^{22}_{\rm
inst})_{\rm Newt}&=\frac{2}{\left(1-e_t \cos u\right){}^2} \biggl\{
1-e_t^2-\frac{1}{2} \left(e_t \cos u\right)+\frac{1}{2} \left(e_t \cos
u\right){}^2 +i \left(e_t \sin u\right) \sqrt{1-e_t^2} \biggr\} ,\\
(\hat{H}^{22}_{\rm inst})_{\rm 1PN}&=-\frac{x}{42 \left(1-e_t^2\right)
\left(1-e_t \cos u\right){}^3} \biggl\{ 214-110 \nu +e_t^2 (64+46 \nu )+e_t^4
(-278+64 \nu ) +\left(e_t \cos u\right) \left(-405+123 \nu \rl+e_t^2 (207-89
\nu )+e_t^4 (114-34 \nu )\right)+\left(e_t \cos u\right){}^2 \left(54+34 \nu
+e_t^2 (114-34 \nu )\right)+\left(e_t \cos u\right){}^3 \left(-27-17 \nu
\rl+e_t^2 (-57+17 \nu )\right)+i \left(e_t \sin u\right)
\sqrt{1-e_t^2}\left(-20-38 \nu +e_t^2 (272-46 \nu)+\left(e_t \cos u\right)
\left(-138+50 \nu +e_t^2 (-114\rrll+34 \nu)\right)\right) \biggr\} ,\\
(\hat{H}^{22}_{\rm inst})_{\rm 2PN}&=\frac{x^2}{3024 \left(1-e_t^2\right){}^2
\left(1-e_t \cos u\right){}^5} \biggl\{ -38932-17836 \nu +8188 \nu ^2+e_t^2
\left(182850-92982 \nu \rl-3966 \nu ^2\right)+e_t^4 \left(-196098+212448 \nu
-30360 \nu ^2\right)+e_t^6 \left(53374-133790 \nu +39866 \nu ^2\right)\nn+e_t^8
\left(-1194+32160 \nu -13728 \nu ^2\right) +\left(e_t \cos u\right)
\left(-4628+84514 \nu -30202 \nu ^2 \rl+e_t^2 \left(-121926-34230 \nu +65946
\nu ^2\right)+e_t^4 \left(199848-80970 \nu -41286 \nu ^2\right)+e_t^6
\left(-35494 \rrll+3470 \nu +5542 \nu ^2\right)\right)+\left(e_t \cos
u\right){}^2 \left(14904+7158 \nu +18558 \nu ^2+e_t^2 \left(-151740+4032 \nu
\rrll-27288 \nu ^2\right)+e_t^4 \left(4320+86670 \nu -1098 \nu ^2\right)+e_t^6
\left(-18684+11004 \nu +9828 \nu ^2\right)\right) \nn+\left(e_t \cos
u\right){}^3 \left(104628-84714 \nu -9786 \nu ^2+e_t^2 \left(140196-58056 \nu
+19248 \nu ^2\right)\rl+e_t^4 \left(-23604-11202 \nu -9138 \nu ^2\right)+e_t^6
\left(5580-9324 \nu -324 \nu ^2\right)\right)+\left(e_t \cos
u\right){}^4\,\times\nn\times \left(-102888+57384 \nu -648 \nu ^2+e_t^2
\left(-59472+70128 \nu +1296 \nu ^2\right)+e_t^4 \left(11160 \rrll-18648 \nu
-648 \nu ^2\right)\right)+\left(e_t \cos u\right){}^5 \left(25722-14346 \nu
+162 \nu ^2+e_t^2 \left(14868-17532 \nu \rrll-324 \nu ^2\right)+e_t^4
\left(-2790+4662 \nu +162 \nu ^2\right)\right)+i \left(e_t \sin u\right)
\left(75600-30240 \nu \rl+e_t^2 (-196560+78624 \nu )+e_t^4 (120960-48384 \nu
)+\left(e_t \cos u\right) \left(-136080+54432 \nu \rrll+e_t^2 (408240-163296
\nu )+e_t^4 (-272160+108864 \nu )\right)+\left(e_t \cos u\right){}^2
\left(45360-18144 \nu \rrll+e_t^2 (-226800+90720 \nu )+e_t^4 (181440-72576 \nu
)\right)+\left(e_t \cos u\right){}^3 \left(15120-6048 \nu \rrll+e_t^2
(15120-6048 \nu )+e_t^4 (-30240+12096 \nu )\right)\right)+\sqrt{1-e_t^2}
\left(30240-12096 \nu \rl+e_t^2 (-151200+60480 \nu )+e_t^4 (120960-48384 \nu
)+\left(e_t \cos u\right) \left(-37800+15120 \nu \rrll+e_t^2 (309960-123984 \nu
)+e_t^4 (-272160+108864 \nu )\right)+\left(e_t \cos u\right){}^2
\left(37800-15120 \nu \rrll+e_t^2 (-219240+87696 \nu )+e_t^4 (181440-72576 \nu
)\right)+\left(e_t \cos u\right){}^3 \left(-98280+39312 \nu \rrll+e_t^2
(128520-51408 \nu )+e_t^4 (-30240+12096 \nu )\right)+\left(e_t \cos
u\right){}^4 \left(83160-33264 \nu \rrll+e_t^2 (-83160+33264 \nu
)\right)+\left(e_t \cos u\right){}^5 \left(-15120+6048 \nu +e_t^2 (15120-6048
\nu)\right) \rl+i \left(e_t \sin u\right) \left(-93152+50476 \nu +4844 \nu
^2+e_t^2 \left(105604-154688 \nu -4624 \nu ^2\right) \rrll+e_t^4
\left(-47816+142684 \nu -10612 \nu ^2\right)+e_t^6 \left(5124-44520 \nu +7368
\nu ^2\right)+\left(e_t \cos u\right)\,\times\rrll\times\left(153308-10456 \nu
-9488 \nu ^2+e_t^2 \left(-81832+21920 \nu +25024 \nu ^2\right)+e_t^4
\left(19244 \rrrrllll+6680 \nu -6464 \nu ^2\right)\right)+\left(e_t \cos
u\right){}^2 \left(-82596+18648 \nu -1464 \nu ^2+e_t^2 \left(-26904
\rrrrllll-11664 \nu -2256 \nu ^2\right)+e_t^4 \left(18780-25128 \nu -5352 \nu
^2\right)\right)+\left(e_t \cos u\right){}^3 \left(17316 \rrrlll-14148 \nu
-1260 \nu ^2+e_t^2 \left(18504+10872 \nu +3960 \nu ^2\right)+e_t^4
\left(-5580+9324 \nu \rrrrllll+324 \nu ^2\right)\right)\right)\right) \biggr\}
,\\
(\hat{H}^{22}_{\rm inst})_{\rm 2.5PN}&=\frac{i x^{5/2} \nu }{105 \left(1-e_t
\cos u\right){}^5} \biggl\{ \sqrt{1-e_t^2} \left(-2352+1500 e_t^2+1404
\left(e_t \cos u\right)-552 \left(e_t \cos u\right){}^2\right) \nn+i \left(e_t
\sin u\right) \left(2539-2175 e_t^2+2 \left(e_t \cos u\right)-366 \left(e_t
\cos u\right){}^2\right) \biggr\} ,\\
\label{ellWF:H22-QKR-3PN} (\hat{H}^{22}_{\rm inst})_{\rm
3PN}&=\frac{x^3}{1277337600 \left(1-e_t^2\right){}^3 \left(1-e_t \cos
u\right){}^8} \biggl\{ 186870371328-20826685440 \log \left(1-e_t \cos u\right)
\nn+20826685440 \log \left(\frac{x}{x_0}\right)+15657907200 \nu -13806489600
\nu ^2+2934656000 \nu ^3 \nn+e_t^2 \left(-319951363584+75496734720 \log
\left(1-e_t \cos u\right)-75496734720 \log \left(\frac{x}{x_0}\right) \rl+\nu
\left(-572456762880+18002476800 \pi ^2\right)+128784537600 \nu ^2-2970944000
\nu ^3\right) \nn+e_t^4 \left(7924875264-101530091520 \log \left(1-e_t \cos
u\right)+101530091520 \log \left(\frac{x}{x_0}\right) \rl+\nu
\left(1455622894080-43369603200 \pi ^2\right)-240132902400 \nu ^2+5117260800
\nu ^3\right) \nn+e_t^6 \left(292424159232+59876720640 \log \left(1-e_t \cos
u\right)-59876720640 \log \left(\frac{x}{x_0}\right) \rl+\nu
\left(-1579426229760+34368364800 \pi ^2\right)+359519040000 \nu ^2-32956160000
\nu ^3\right) \nn+e_t^8 \left(-171895234560-13016678400 \log \left(1-e_t \cos
u\right)+13016678400 \log \left(\frac{x}{x_0}\right) \rl+\nu
\left(873551592960-9001238400 \pi ^2\right)-368015424000 \nu ^2+53854515200 \nu
^3\right) \nn+e_t^{10} \left(4469829120-197763148800 \nu +146426457600 \nu
^2-32062809600 \nu ^3\right) \nn+e_t^{12} \left(157363200+4813747200 \nu
-12775219200 \nu ^2+6083481600 \nu ^3\right) \nn+\left(e_t \cos u\right)
\left(-815811686400+65083392000 \log \left(1-e_t \cos
u\right)-65083392000\,\times\rl\times\log \left(\frac{x}{x_0}\right)+\nu
\left(-107770229760-2454883200 \pi ^2\right)+70743936000 \nu ^2 \rl-21185804800
\nu ^3+e_t^2 \left(1636480788480-234300211200 \log \left(1-e_t \cos u\right)
\rrll+234300211200 \log \left(\frac{x}{x_0}\right)+\nu
\left(1487299445760-49915958400 \pi ^2\right) \rrll-420029798400 \nu
^2+30425203200 \nu ^3\right)+e_t^4 \left(-691922419200+312400281600
\,\times\rrll\times \log \left(1-e_t \cos u\right)-312400281600 \log
\left(\frac{x}{x_0}\right)+\nu  \left(-3463303150080 \rrrlll+135836870400 \pi
^2\right)+230527872000 \nu ^2+22174886400 \nu ^3\right)+e_t^6
\left(-529844974080 \rrll-182233497600 \log \left(1-e_t \cos
u\right)+182233497600 \log \left(\frac{x}{x_0}\right)+\nu \left(3240401410560
\rrrlll-106378272000 \pi ^2\right)-90015513600 \nu ^2-31134348800 \nu
^3\right)+e_t^8 \left(500039447040 \rrll+39050035200 \log \left(1-e_t \cos
u\right)-39050035200 \log \left(\frac{x}{x_0}\right)+\nu  \left(-1788108142080
\rrrlll+27003715200 \pi ^2\right)+486709862400 \nu ^2-26111232000 \nu
^3\right)+e_t^{10} \left(-10221719040 \rrll+397006502400 \nu -243769420800 \nu
^2+31914777600 \nu ^3\right)+e_t^{12} \left(-157363200 \rrll-4813747200 \nu
+12775219200 \nu ^2-6083481600 \nu ^3\right)\right)+\left(e_t \cos u\right){}^2
\left(1161019760640 \rl-65083392000 \log \left(1-e_t \cos u\right)+65083392000
\log \left(\frac{x}{x_0}\right)+\nu \left(1194968125440 \rrll-14729299200 \pi
^2\right)-260655360000 \nu ^2+53078886400 \nu ^3+e_t^2 \left(-2802117381120
\rrll+234300211200 \log \left(1-e_t \cos u\right)-234300211200 \log
\left(\frac{x}{x_0}\right)+\nu  \left(-2357003520000 \rrrlll+66281846400 \pi
^2\right)+972179174400 \nu ^2-130893849600 \nu ^3\right)+e_t^4
\left(1596715430400 \rrll-312400281600 \log \left(1-e_t \cos
u\right)+312400281600 \log \left(\frac{x}{x_0}\right)+\nu \left(3961710581760
\rrrlll-171023529600 \pi ^2\right)-332844249600 \nu ^2+78350246400 \nu
^3\right)+e_t^6 \left(-86515975680 \rrll+182233497600 \log \left(1-e_t \cos
u\right)-182233497600 \log \left(\frac{x}{x_0}\right)+\nu  \left(-1923370721280
\rrrlll+117834393600 \pi ^2\right)-784967232000 \nu ^2+19523494400 \nu
^3\right)+e_t^8 \left(-495126120960 \rrll-39050035200 \log \left(1-e_t \cos
u\right)+39050035200 \log \left(\frac{x}{x_0}\right)+\nu \left(1028316948480
\rrrlll-27003715200 \pi ^2\right)-47218214400 \nu ^2-15916761600 \nu
^3\right)+e_t^{10} \left(6089771520 \rrll-229606041600 \nu +124910784000 \nu
^2-4142016000 \nu ^3\right)\right)+\left(e_t \cos u\right){}^3
\left(-334437120000 \rl+13016678400 \log \left(1-e_t \cos u\right)-13016678400
\log \left(\frac{x}{x_0}\right)+\nu \left(-3817315046400 \rrll+84284323200 \pi
^2\right)+457420492800 \nu ^2-48936947200 \nu ^3+e_t^2 \left(2153965992960
\rrll-52066713600 \log \left(1-e_t \cos u\right)+52066713600 \log
\left(\frac{x}{x_0}\right)+\nu  \left(2622913935360 \rrrlll-87557500800 \pi
^2\right)-969149184000 \nu ^2+146756108800 \nu ^3\right)+e_t^4
\left(-713296212480 \rrll+78100070400 \log \left(1-e_t \cos
u\right)-78100070400 \log \left(\frac{x}{x_0}\right)+\nu \left(-3996933972480
\rrrlll+139110048000 \pi ^2\right)+791266867200 \nu ^2-150936691200 \nu
^3\right)+e_t^6 \left(564367726080 \rrll-52066713600 \log \left(1-e_t \cos
u\right)+52066713600 \log \left(\frac{x}{x_0}\right)+\nu  \left(239607559680
\rrrlll-58917196800 \pi ^2\right)+810237081600 \nu ^2+61642892800 \nu
^3\right)+e_t^8 \left(189541040640 \rrll+13016678400 \log \left(1-e_t \cos
u\right)-13016678400 \log \left(\frac{x}{x_0}\right)+\nu \left(-103681282560
\rrrlll+9001238400 \pi ^2\right)-76422144000 \nu ^2-12815411200 \nu
^3\right)+e_t^{10} \left(-337881600 \rrll+30362688000 \nu -27567820800 \nu
^2+4290048000 \nu ^3\right)\right)+\left(e_t \cos u\right){}^4
\left(-687858478080 \rl+13016678400 \log \left(1-e_t \cos u\right)-13016678400
\log \left(\frac{x}{x_0}\right)+\nu \left(5807370984960 \rrll-153021052800 \pi
^2\right)-455413900800 \nu ^2+17977395200 \nu ^3+e_t^2 \left(-1490127552000
\rrll-39050035200 \log \left(1-e_t \cos u\right)+39050035200 \log
\left(\frac{x}{x_0}\right)+\nu  \left(-384485191680 \rrrlll+90830678400 \pi
^2\right)-203492160000 \nu ^2-53794060800 \nu ^3\right)+e_t^4
\left(-664152491520 \rrll+39050035200 \log \left(1-e_t \cos
u\right)-39050035200 \log \left(\frac{x}{x_0}\right)+\nu \left(2723950126080
\rrrlll-99013622400 \pi ^2\right)-465812467200 \nu ^2+53517811200 \nu
^3\right)+e_t^6 \left(-224123312640 \rrll-13016678400 \log \left(1-e_t \cos
u\right)+13016678400 \log \left(\frac{x}{x_0}\right)+\nu  \left(255670333440
\rrrlll+18002476800 \pi ^2\right)-514787366400 \nu ^2-17563020800 \nu
^3\right)+e_t^8 \left(-33410741760 \rrll-27429388800 \nu -3469593600 \nu
^2-138124800 \nu ^3\right)\right)+\left(e_t \cos u\right){}^5
\left(787721912832 \rl-5206671360 \log \left(1-e_t \cos u\right)+5206671360
\log \left(\frac{x}{x_0}\right)+\nu \left(-5142719623680 \rrll+140746636800 \pi
^2\right)+348355737600 \nu ^2-7195404800 \nu ^3+e_t^2 \left(1833256917504
\rrll+15620014080 \log \left(1-e_t \cos u\right)-15620014080 \log
\left(\frac{x}{x_0}\right)+\nu  \left(-2284122493440 \rrrlll-38459836800 \pi
^2\right)+1005667968000 \nu ^2+20515392000 \nu ^3\right)+e_t^4
\left(477180919296 \rrll-15620014080 \log \left(1-e_t \cos u\right)+15620014080
\log \left(\frac{x}{x_0}\right)+\nu \left(-597097059840 \rrrlll+45824486400 \pi
^2\right)-7296998400 \nu ^2-18373747200 \nu ^3\right)+e_t^6 \left(-10724639232
\rrll+5206671360 \log \left(1-e_t \cos u\right)-5206671360 \log
\left(\frac{x}{x_0}\right)+\nu  \left(-375209879040 \rrrlll-4909766400 \pi
^2\right)+288255360000 \nu ^2+3982937600 \nu ^3\right)+e_t^8 \left(12237465600
\rrll+24072192000 \nu +7993420800 \nu ^2+1070822400 \nu
^3\right)\right)+\left(e_t \cos u\right){}^6 \left(-411239646720 \rl+\nu
\left(2829529175040-74464790400 \pi ^2\right)-207633139200 \nu ^2+3495795200
\nu ^3 \rl+e_t^2 \left(-1455591651840+\nu  \left(2185925806080-4091472000 \pi
^2\right)-724992192000 \nu ^2 \rrll-10543577600 \nu ^3\right)+e_t^4
\left(9839024640+\nu  \left(-146180229120-7364649600 \pi ^2\right)
\rrll+15925593600 \nu ^2+10655961600 \nu ^3\right)+e_t^6
\left(-1425415680+162493286400 \nu \rrll-69507648000 \nu ^2-3664371200 \nu
^3\right)+e_t^8 \left(-1385856000-6721920000 \nu \rrll+422092800 \nu
^2+56192000 \nu ^3\right)\right)+\left(e_t \cos u\right){}^7
\left(132690700800+\nu  \left(-909674841600 \rrll+22912243200 \pi
^2\right)+71153510400 \nu ^2-196672000 \nu ^3+e_t^2 \left(518098291200
\rrll+\nu  \left(-814416422400+5728060800 \pi ^2\right)+246203596800 \nu
^2+590016000 \nu ^3\right) \rl+e_t^4 \left(-26003980800+72602611200 \nu
+9760665600 \nu ^2-590016000 \nu ^3\right) \rl+e_t^6
\left(-4850496000-23526720000 \nu +1477324800 \nu ^2+196672000 \nu
^3\right)\right) \nn+\left(e_t \cos u\right){}^8 \left(-18955814400+\nu
\left(129953548800-3273177600 \pi ^2\right)-10164787200 \nu ^2 \rl+28096000 \nu
^3+e_t^2 \left(-74014041600+\nu  \left(116345203200-818294400 \pi ^2\right)
\rrll-35171942400 \nu ^2-84288000 \nu ^3\right)+e_t^4
\left(3714854400-10371801600 \nu \rrll-1394380800 \nu ^2+84288000 \nu
^3\right)+e_t^6 \left(692928000+3360960000 \nu -211046400 \nu ^2 \rrll-28096000
\nu ^3\right)\right)+i \left(e_t \sin u\right) \left(74511360000+\nu
\left(-225308160000+2727648000 \pi ^2\right) \rl+13199155200 \nu ^2+e_t^2
\left(-387763200000+\nu \left(693492940800-7091884800 \pi ^2\right)
\rrll-37407744000 \nu ^2\right)+e_t^4 \left(417871872000+\nu
\left(-487101542400+4364236800 \pi ^2\right) \rrll+729907200 \nu
^2\right)+e_t^6 \left(-104620032000+18916761600 \nu +23478681600 \nu ^2\right)
\rl+\left(e_t \cos u\right) \left(-241173504000+\nu
\left(1018828800000-13092710400 \pi ^2\right)-56932761600 \nu ^2 \rrll+e_t^2
\left(1711176192000+\nu  \left(-3424055500800+36004953600 \pi
^2\right)+194885222400 \nu ^2\right) \rrll+e_t^4 \left(-1894109184000+\nu
\left(2447865446400-22912243200 \pi ^2\right)-12043468800 \nu ^2\right)
\rrll+e_t^6 \left(424106496000-42638745600 \nu -125908992000 \nu
^2\right)\right)+\left(e_t \cos u\right){}^2 \left(196314624000 \rrll+\nu
\left(-1765980057600+24548832000 \pi ^2\right)+89961062400 \nu ^2+e_t^2
\left(-3063785472000 \rrrlll+\nu  \left(6872258764800-73646496000 \pi
^2\right)-423163699200 \nu ^2\right)+e_t^4 \left(3526820352000 \rrrlll+\nu
\left(-5087179468800+49097664000 \pi ^2\right)+59122483200 \nu ^2\right)+e_t^6
\left(-659349504000 \rrrlll-19099238400 \nu +274080153600 \nu
^2\right)\right)+\left(e_t \cos u\right){}^3 \left(136249344000+\nu
\left(1381795430400 \rrrlll-21821184000 \pi ^2\right)-54499737600 \nu ^2+e_t^2
\left(2898643968000+\nu  \left(-7118338867200 \rrrrllll+76374144000 \pi
^2\right)+491957452800 \nu ^2\right)+e_t^4 \left(-3523627008000+\nu
\left(5612094259200 \rrrrllll-54552960000 \pi	^2\right)-130653388800 \nu
^2\right)\right.\right.
\left.+e_t^6 \left(488733696000 +124449177600 \nu \rrll-306804326400 \nu
^2\right)\right) +\left(e_t \cos u\right){}^4 \left(-292571136000+\nu
\left(-358171545600 \rrll+8182944000 \pi ^2\right) -4926873600 \nu ^2+e_t^2
\left(-1608532992000+\nu \left(3959624908800 \rrrlll-40914720000 \pi
^2\right)-324626227200 \nu ^2\right)+e_t^4 \left(2082972672000+\nu
\left(-3485337292800 \rrrlll+32731776000 \pi ^2\right)+146711347200 \nu
^2\right)+e_t^6 \left(-181868544000-116116070400 \nu \rrll+182841753600 \nu
^2\right)\right)
+\left(e_t \cos u\right){}^5 \left(149935104000-107418009600 \nu +18977587200
\nu ^2 \rl+e_t^2 \left(546974208000+\nu  \left(-1093948416000+9819532800 \pi
^2\right)+116055244800 \nu ^2\right) \rl+e_t^4 \left(-739031040000+\nu
\left(1166391705600-9819532800 \pi ^2\right)-82114560000 \nu ^2\right)
\rl+e_t^6 \left(42121728000+34974720000 \nu -52918272000 \nu ^2\right)\right)
+\left(e_t \cos u\right){}^6 \left(-23265792000 \rl+\nu
\left(56253542400-545529600 \pi ^2\right)-5778432000 \nu ^2+e_t^2
\left(-96712704000 \rrll+\nu  \left(110966169600-545529600 \pi
^2\right)-17700249600 \nu ^2\right)+e_t^4 \left(129102336000 \rrll+\nu
\left(-166733107200+1091059200 \pi ^2\right)+18247680000 \nu ^2\right)+e_t^6
\left(-9123840000 \rrll\left.-486604800 \nu +5231001600 \nu ^2\right)\right)
\right)
+\sqrt{1-e_t^2} \left( -39536640000+\nu \left(-37022515200 \rrll+1091059200 \pi
^2\right)-4866048000 \nu ^2+e_t^2 \left(-183389184000+\nu  \left(472087756800
\rrrlll-5455296000 \pi ^2\right)-25911705600 \nu ^2\right)+e_t^4
\left(332107776000+\nu \left(-469492531200 \rrrlll+4364236800 \pi
^2\right)+12773376000 \nu ^2\right)+e_t^6 \left(-109181952000+34427289600 \nu
\rrll+18004377600 \nu ^2\right)+\left(e_t \cos u\right) \left(285956352000+\nu
\left(94893004800-4637001600 \pi ^2\right) \rrll+26793676800 \nu ^2+e_t^2
\left(809740800000+\nu  \left(-2359414886400+27549244800 \pi ^2\right)
\rrrlll+142483968000 \nu ^2\right)+e_t^4 \left(-1500187392000+\nu
\left(2355425740800-22912243200 \pi ^2\right) \rrrlll-65813299200 \nu
^2\right)+e_t^6 \left(442810368000-106231910400 \nu -103464345600 \nu
^2\right)\right) \rl+\left(e_t \cos u\right){}^2 \left(-702383616000+\nu
\left(-158623027200+8728473600 \pi ^2\right)-32176742400 \nu ^2 \rrll+e_t^2
\left(-1736266752000+\nu \left(5106816000000-57826137600 \pi
^2\right)-358262784000 \nu ^2\right) \rrll+e_t^4 \left(2858955264000+\nu
\left(-4921064755200+49097664000 \pi ^2\right)+151394918400 \nu ^2\right)
\rrll+e_t^6 \left(-688545792000+80168140800 \nu +239044608000 \nu
^2\right)\right)+\left(e_t \cos u\right){}^3 \left(714016512000 \rrll+\nu
\left(650230732800-12819945600 \pi ^2\right)-68824166400 \nu ^2+e_t^2
\left(2664617472000 \rrrlll+\nu  \left(-6534899712000+67372905600 \pi
^2\right)+540040089600 \nu ^2\right)+e_t^4 \left(-3083629824000 \rrrlll+\nu
\left(5509679155200-54552960000 \pi ^2\right)-189593395200 \nu ^2\right)+e_t^6
\left(509718528000 \rrrlll+53100748800 \nu -281622528000 \nu
^2\right)\right)+\left(e_t \cos u\right){}^4 \left(-163012608000 \rrll+\nu
\left(-1666905292800+19093536000 \pi ^2\right)+237828096000 \nu ^2+e_t^2
\left(-3119440896000 \rrrlll+\nu \left(5801890406400-51825312000 \pi
^2\right)-524012544000 \nu ^2\right)+e_t^4 \left(2129504256000 \rrrlll+\nu
\left(-3504101990400+32731776000 \pi ^2\right)+111006720000 \nu ^2\right)+e_t^6
\left(-188255232000 \rrrlll-94401331200 \nu +175177728000 \nu
^2\right)\right)+\left(e_t \cos u\right){}^5 \left(-234862848000 \rrll+\nu
\left(2089780070400-21548419200 \pi ^2\right)-284815872000 \nu ^2+e_t^2
\left(2485790208000 \rrrlll+\nu \left(-3852277862400+31367952000 \pi
^2\right)+325021593600 \nu ^2\right)+e_t^4 \left(-952300800000 \rrrlll+\nu
\left(1192592332800-9819532800 \pi ^2\right)+12165120000 \nu ^2\right)+e_t^6
\left(42577920000 \rrrlll+33423667200 \nu -52370841600 \nu
^2\right)\right)+\left(e_t \cos u\right){}^6 \left(180195840000+\nu
\left(-1346993049600 \rrrlll+14183769600 \pi ^2\right)+171953971200 \nu
^2+e_t^2 \left(-1212558336000+\nu  \left(1821503692800 \rrrrllll-15274828800
\pi ^2\right)-124084224000 \nu ^2\right)+e_t^4 \left(236763648000+\nu
\left(-152134963200 \rrrrllll+1091059200 \pi ^2\right)-53100748800 \nu
^2\right)+e_t^6 \left(-9123840000-486604800 \nu +5231001600 \nu
^2\right)\right) \rl+\left(e_t \cos u\right){}^7 \left(-44934912000+\nu
\left(424780646400-4637001600 \pi ^2\right)-52218777600 \nu ^2 \rrll+e_t^2
\left(338950656000+\nu  \left(-520930713600+4637001600 \pi
^2\right)+28435968000 \nu ^2\right) \rrll+e_t^4 \left(-25774848000-11146291200
\nu +23782809600 \nu ^2\right)\right)+\left(e_t \cos u\right){}^8
\left(4561920000 \rrll+\nu \left(-50140569600+545529600 \pi
^2\right)+6325862400 \nu ^2+e_t^2 \left(-47443968000 \rrrlll+\nu
\left(65225318400-545529600 \pi ^2\right)-3710361600 \nu ^2\right)+e_t^4
\left(4561920000+243302400 \nu \rrrlll-2615500800 \nu ^2\right)\right)
+i \left(e_t \sin u\right) \left( -328593346560+26033356800 \log \left(1-e_t
\cos u\right) \rrll-26033356800 \log \left(\frac{x}{x_0}\right)+71427240960 \nu
-11877196800 \nu ^2+2718080000 \nu ^3 \rrll+e_t^2
\left(962564305920-78100070400 \log \left(1-e_t \cos u\right)+78100070400 \log
\left(\frac{x}{x_0}\right) \rrrlll+\nu \left(-85968921600-3273177600 \pi
^2\right)+106786867200 \nu ^2-6687820800 \nu ^3\right) \rrll+e_t^4
\left(-1161059174400+78100070400 \log \left(1-e_t \cos u\right)-78100070400
\log \left(\frac{x}{x_0}\right) \rrrlll+\nu  \left(455668669440+4909766400 \pi
^2\right)-231924172800 \nu ^2+11614080000 \nu ^3\right) \rrll+e_t^6
\left(483224340480-26033356800 \log \left(1-e_t \cos u\right)+26033356800 \log
\left(\frac{x}{x_0}\right) \rrrlll+\nu \left(-538960911360-3273177600 \pi
^2\right)+319443148800 \nu ^2-12469836800 \nu ^3\right) \rrll+e_t^8
\left(-22426122240+229909708800 \nu -173743948800 \nu ^2+7364044800 \nu
^3\right) \rrll+e_t^{10} \left(294220800-9305395200 \nu +23887411200 \nu
^2-1261209600 \nu ^3\right) +\left(e_t \cos u\right)\times\rrll\times
\left(925176069120-78100070400 \log \left(1-e_t \cos u\right)+78100070400 \log
\left(\frac{x}{x_0}\right) \rrrlll+\nu  \left(-458849817600+8182944000 \pi
^2\right)-11032934400 \nu ^2-9615590400 \nu ^3 \rrrlll+e_t^2
\left(-2356066314240+234300211200 \log \left(1-e_t \cos u\right)-234300211200
\log \left(\frac{x}{x_0}\right) \rrrrllll-300753884160 \nu -62224051200 \nu
^2+13476403200 \nu ^3\right)+e_t^4 \left(3159065625600 \rrrrllll-234300211200
\log \left(1-e_t \cos u\right)+234300211200 \log \left(\frac{x}{x_0}\right)+\nu
\left(-338253788160 \rrrrrlllll-8182944000 \pi ^2\right)-32412595200 \nu
^2-11630284800 \nu ^3\right)+e_t^6 \left(-1380322575360 \rrrrllll+78100070400
\log \left(1-e_t \cos u\right)-78100070400 \log \left(\frac{x}{x_0}\right)+\nu
\left(811123860480 \rrrrrlllll+9819532800 \pi ^2\right)-319331097600 \nu
^2+5767065600 \nu ^3\right)+e_t^8 \left(48416071680 \rrrrllll-459194112000 \nu
+253455436800 \nu ^2-6922828800 \nu ^3\right)+e_t^{10} \left(-294220800
\rrrrllll+9305395200 \nu -23887411200 \nu ^2+1261209600 \nu ^3\right)\right)
+\left(e_t \cos u\right){}^2 \left(-901804861440 \rrrlll+78100070400 \log
\left(1-e_t \cos u\right)-78100070400 \log \left(\frac{x}{x_0}\right)+\nu
\left(1567758735360 \rrrrllll-34368364800 \pi ^2\right)-4901222400 \nu
^2+16293504000 \nu ^3+e_t^2 \left(1178660367360 \rrrrllll-234300211200 \log
\left(1-e_t \cos u\right)+234300211200 \log \left(\frac{x}{x_0}\right)+\nu
\left(638531481600 \rrrrrlllll+26185420800 \pi ^2\right)+85797964800 \nu
^2-23253580800 \nu ^3\right)+e_t^4 \left(-2558005079040 \rrrrllll+234300211200
\log \left(1-e_t \cos u\right)-234300211200 \log \left(\frac{x}{x_0}\right)+\nu
\left(-627219118080 \rrrrrlllll-6546355200 \pi ^2\right)+636385766400 \nu
^2+18039628800 \nu ^3\right)+e_t^6 \left(1322237813760 \rrrrllll-78100070400
\log \left(1-e_t \cos u\right)+78100070400 \log \left(\frac{x}{x_0}\right)+\nu
\left(-12701306880 \rrrrrlllll-9819532800 \pi ^2\right)-123584793600 \nu
^2+10020940800 \nu ^3\right)+e_t^8 \left(-31024880640 \rrrrllll+275186073600
\nu -105116083200 \nu ^2-1940428800 \nu ^3\right)\right) +\left(e_t \cos
u\right){}^3 \left(452697154560 \rrrlll-26033356800 \log \left(1-e_t \cos
u\right)+26033356800 \log \left(\frac{x}{x_0}\right)+\nu \left(-2649367956480
\rrrrllll+58917196800 \pi ^2\right)+75336499200 \nu ^2-10847257600 \nu ^3+e_t^2
\left(1093220843520 \rrrrllll+78100070400 \log \left(1-e_t \cos
u\right)-78100070400 \log \left(\frac{x}{x_0}\right)+\nu  \left(5716638720
\rrrrrlllll-49097664000 \pi ^2\right)-361501670400 \nu ^2+20455526400 \nu
^3\right)+e_t^4 \left(211920660480 \rrrrllll-78100070400 \log \left(1-e_t \cos
u\right)+78100070400 \log \left(\frac{x}{x_0}\right)+\nu \left(602037273600
\rrrrrlllll+19639065600 \pi ^2\right)-554518809600 \nu ^2-36066355200 \nu
^3\right)+e_t^6 \left(-442958069760 \rrrrllll+26033356800 \log \left(1-e_t \cos
u\right)-26033356800 \log \left(\frac{x}{x_0}\right)+\nu  \left(-367892106240
\rrrrrlllll+3273177600 \pi ^2\right)+163837209600 \nu ^2-587878400 \nu
^3\right)+e_t^8 \left(5034931200 \rrrrllll-45901670400 \nu +25404595200 \nu
^2+1499212800 \nu ^3\right)\right) +\left(e_t \cos u\right){}^4
\left(-287513978880 \rrrlll+\nu  \left(2242649057280-52370841600 \pi
^2\right)-56454220800 \nu ^2-971289600 \nu ^3 \rrrlll+e_t^2
\left(-1158390312960+\nu  \left(-333825269760+39278131200 \pi
^2\right)+302949196800 \nu ^2 \rrrrllll+1949107200 \nu ^3\right)+e_t^4
\left(427268290560+\nu  \left(-205133137920-11456121600 \pi ^2\right)
\rrrrllll+280446028800 \nu ^2+20456678400 \nu ^3\right)+e_t^6
\left(28699361280+137865216000 \nu \rrrrllll-38359372800 \nu ^2-2274432000 \nu
^3\right)\right)+\left(e_t \cos u\right){}^5 \left(169541452800 \rrrlll+\nu
\left(-933238149120+24548832000 \pi ^2\right)+9044275200 \nu ^2+2341017600 \nu
^3 \rrrlll+e_t^2 \left(328893465600+\nu \left(69430671360-16365888000 \pi
^2\right)-74749132800 \nu ^2 \rrrrllll-5811609600 \nu ^3\right)+e_t^4
\left(-90193536000+\nu \left(163341803520+1636588800 \pi ^2\right)
\rrrrllll-128144793600 \nu ^2-3793766400 \nu ^3\right)+e_t^6
\left(-12266726400-36156672000 \nu \rrrrllll-1583001600 \nu ^2-399667200 \nu
^3\right)\right)+\left(e_t \cos u\right){}^6 \left(-29502489600+\nu
\left(159620889600 \rrrrllll-4909766400 \pi ^2\right)-115200000 \nu ^2+81536000
\nu ^3+e_t^2 \left(-48882355200+\nu  \left(6869283840 \rrrrrlllll+3273177600
\pi ^2\right)+2940825600 \nu ^2-128025600 \nu ^3\right)+e_t^4
\left(11003212800-50441702400 \nu \rrrrllll+30168576000 \nu ^2+1380019200 \nu
^3\right)+e_t^6 \left(1385856000+6721920000 \nu -422092800 \nu ^2
\rrrrllll-56192000 \nu ^3\right)\right) \right) \right) \biggr\}.  \end{align}
\ese \end{widetext}

\bibliography{}
\end{document}